\documentclass[%
 reprint,
 amsmath,amssymb,
 aps,
]{revtex4-2}
\usepackage{comment}
\usepackage[normalem]{ulem}
\usepackage{graphicx}
\usepackage{dcolumn}
\usepackage{float}
\usepackage{bm}
\usepackage{hyperref}
\usepackage{dirtytalk}
\usepackage{xcolor}
\usepackage{soul}
\usepackage{amssymb}
\usepackage{comment}
\usepackage{graphicx}
\usepackage{dcolumn}
\usepackage{bm}
\usepackage{hyperref}
\renewcommand{\thefigure}{\textbf{\arabic{figure}}}
\usepackage{dirtytalk}
\hypersetup{
    colorlinks=true,
    linkcolor=blue,
    filecolor=magenta,      
    urlcolor=cyan,
    pdftitle={Overleaf Example},
    pdfpagemode=FullScreen,
    }

\begin{document}
\preprint{APS/123-QED}

\title{Engineering ultra-strong electron-phonon coupling and nonclassical electron transport in crystalline gold with nanoscale interfaces}

\author{Shreya Kumbhakar$^{1,\star,\dagger}$}
\author{Tuhin Kumar Maji$^{1,\star,\dagger}$}%
  \author{Binita Tongbram$^1$}%
   \author{Shinjan Mandal$^1$}
   \author{Shri Hari Soundararaj$^{1,3}$}
   \author{Banashree Debnath$^1$}
  \author{T. Phanindra Sai$^1$}%
 \author{Manish Jain$^1$}
 \author{H. R. Krishnamurthy$^{1,4}$}
  \author{Anshu Pandey$^2$}%
 \author{Arindam Ghosh$^{1,\dagger}$}%
\affiliation{%
 $^1$Department of Physics, Indian Institute of Science, Bangalore 560012, India
}%
\affiliation{%
 $^2$Solid State and Structural Chemistry Unit, Indian Institute of Science, Bangalore 560012, India\affiliation{}
}
\affiliation{$^3$Mechanical Engineering Deptartement, University of California, Riverside}
\affiliation{$^4$International Centre for Theoretical Sciences, Tata Institute of Fundamental Research, Bangalore 560012, India}
\maketitle
{\bf Electrical resistivity in good metals, particularly noble metals such as gold (Au), silver (Ag), or copper, increases linearly with temperature ($T$) for $T > \Theta_{\mathrm{D}}$, where $\Theta_\mathrm{{D}}$ is the Debye temperature~\cite{PBAllen, ziman1972principles}. This is because the coupling ($\lambda$) between the electrons and the lattice vibrations, or phonons, in these metals is rather weak with $\lambda \sim 0.1-0.2$ \cite{PBAllen}, and a perturbative analysis suffices to explain the $T$-linear electron-phonon scattering rate. In this work, we outline a new nanostructuring strategy of crystalline Au where this foundational concept of metallic transport breaks down. We show that by embedding a distributed network of ultra-small Ag nanoparticles (AgNPs) of radius $\sim1-2$ nm inside a crystalline Au shell, an unprecedented enhancement in the electron-phonon interaction, with  $\lambda$ as high as $\approx 20$, can be achieved. This is over hundred times that of bare Au or Ag, and ten times larger than any known metal. With increasing AgNP density, the electrical resistivity deviates from $T$-linearity, and approaches a saturation to the Mott-Ioffe-Regel scale \cite{ioffe1960non}  $\rho_{\mathrm{MIR}}\sim ha/e^2$ for both disorder  ($T\rightarrow 0$) and phonon  ($T \gg \Theta_{\mathrm{D}}$)-dependent components of resistivity (here, $a=0.3$~nm, is the lattice constant of Au). This giant electron-phonon interaction, which we suggest arises from the coulomb interaction-induced coupling of conduction electrons to the localized phonon modes at the buried Au-Ag hetero-interfaces, allows experimental access to a regime of nonclassical metallic transport that has never been probed before.}

The phenomenon of resistivity saturation in disordered metals, especially transition metals and compounds that are usually good superconductors at low temperatures, has remained an open problem in solid-state physics for over 50 years \cite{FISK1973277,PhysRevLett.38.782,ALLEN1980291,gunnarsson2003colloquium,hussey2004universality,PhysRevLett.87.266601,PhysRevB.54.5389,werman2016mott,werman2017non}. It is commonly agreed that saturation occurs when the scattering of electrons by either defects or thermal excitations (phonons) shrinks the mean free path to the order of inter-atomic spacing – the so-called Mott-Ioffe-Regel (MIR) limit \cite{ioffe1960non,gunnarsson2003colloquium}. The quasi-particles become incoherent, resulting in a class of ‘bad metals’. Although a comprehensive theoretical understanding of the saturation remains elusive \cite{PhysRevLett.87.266601,PhysRevB.54.5389,werman2016mott,werman2017non}, the electron-phonon coupling (EPC) seems to play a key role, and resistivity saturation has been linked to, for example, the breakdown of Born-Oppenheimer approximation \textcolor{black}{and delocalization of electronic states} \cite{PhysRevB.68.033103}, intermediate coupling of phonons to local electronic levels or hopping integrals \cite{PhysRevLett.87.266601,PhysRevB.54.5389} or phonon-driven parallel channels of electrical conduction \cite{werman2016mott,werman2017non}. In fact, the limit of extremely strong EPC remains poorly understood in the context of metallic transport in general, and it is not clear though whether the scattering rate of electrons would be limited by, for example, possible universal (Planckian) bound \cite{doi:10.1126/science.1227612,PhysRevLett.123.066601}, polaronic deformation \cite{werman2017non}, or indeed, the stability of the metallic state itself against polaronic self-trapping \cite{doi:10.1073/pnas.2216241120}. The lack of understanding is partly caused by the fact that in most naturally occurring or synthesized metallic solids/alloys so far, the EPC parameter ($\lambda$) has not been found to exceed $\sim 2$ \cite{PBAllen}, let alone a systematic tunability of $\lambda$ over a broad range in the same system. Controlled incorporation of disorder has been shown to assist saturation of resistivity at high temperature, leading to the so called ‘Mooij correlation’ \cite{mooij1973electrical}, but the EPC remains largely intrinsic and $\lesssim 1$, and the fate of metallic transport in the limit $\lambda\gg1$ remains experimentally unknown.

\begin{figure*}
\centering
  \includegraphics[height=9.5cm]{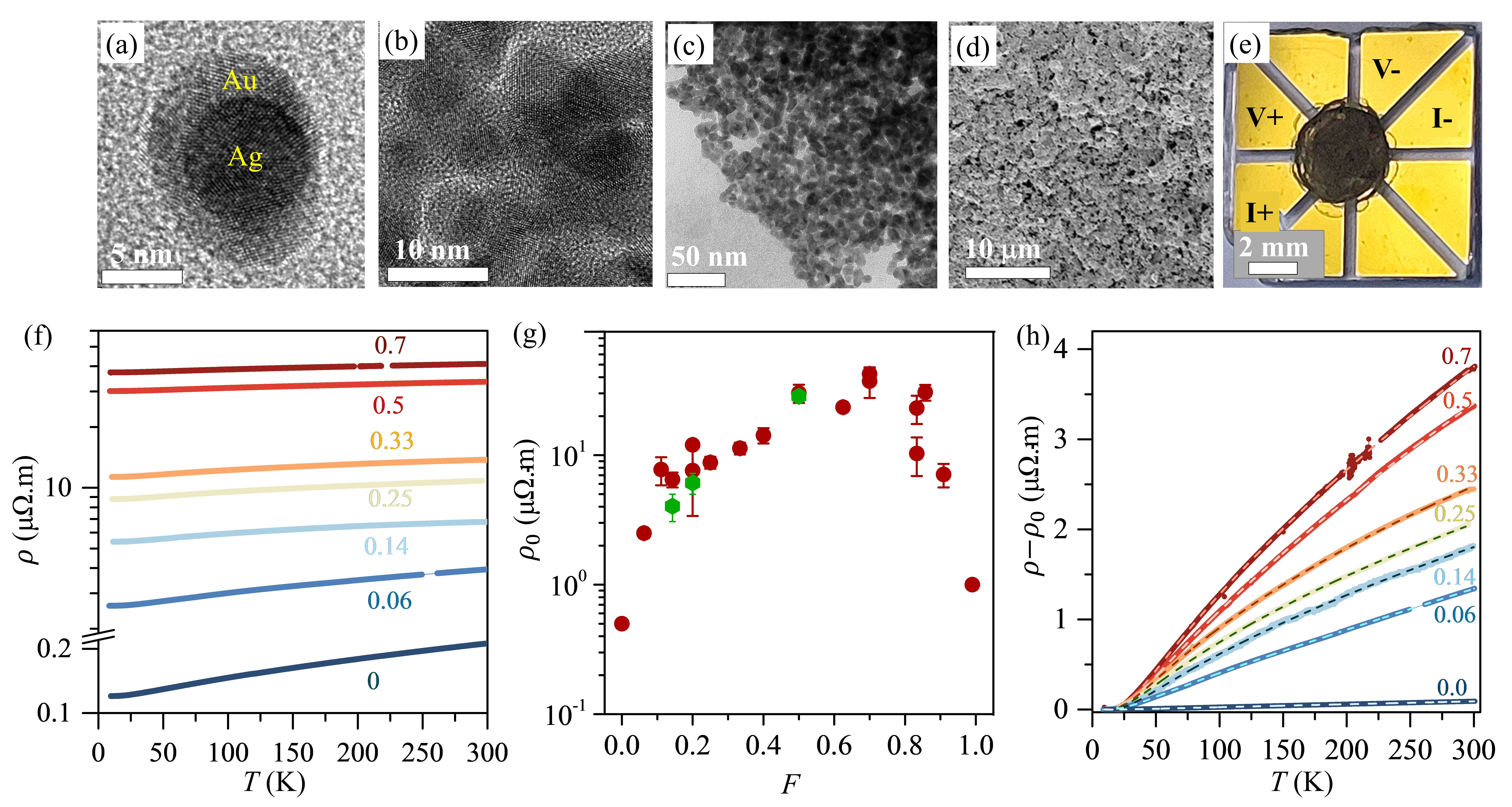}
 \caption{\textbf{Structure and electrical transport:}(a)-(e) Ag@Au nanohybrids at different length scales. (a) High-Resolution Transmission Electron Microscopy (HRTEM) image of single AgNP of high crystallinity is embedded inside a crystalline matrix of Au. The AgNPs are typically spherical and the interface between AgNPs and the Au host is sharp even at the atomic scale. (b)  The interface sharpness remains robust even after forming a dense AgNP network within Au.  (c) A snapshot of the intermediate stage of fusing (or, cross-linking) of the Ag@Au nanohybrids that eventually form a continuous and compact network upon multiple loading via dropcast. (d) Scanning Electron Microscope (SEM) image of a Ag@Au film after multiple (about ten) iterations of dropcast and cross-linking. (e) Optical image of typical Ag@Au film fabricated on pre-patterned Van der Pauw leads. $I+,I-$ and $V+,V-$ represent the current and volatge contacts for four-probe resistivity ($\rho$) measurements. (f) Variation of $\rho$ with temperature ($T$) for films with different Ag volume fraction $F$ (equivalent to AgNP density), showing metallic transport down to the lowest temperature. (g) Residual resistivity ($\rho_0$), defined as the value of $\rho$ observed at $T\sim 6$~K as a function of $F$. \textcolor{black}{The red and green points were evaluated from films in Van der Pauw and Hall bar geometries, respectively. (See Extended Data Fig.~1)} (h) Variation of the $T$-dependent component of $\rho$, obtained by subtracting $\rho_0$ from $\rho$, reveals the emergence of sub-linear behavior in $\rho$ at high temperatures with increasing $F$. The pure AuNP film is represented by $F=0$. Dashed lines represent fit to the data using the two-component parallel channel model given by Eq.~[\ref{fit_RT}].}
  \label{fig1}
\end{figure*}

Well known methods to engineer EPC in solids often depend on confinement or localization of both electrons and phonons, enabled by defects, topological disorder or interfaces that cause acoustic impedance mismatch, lifting of structural symmetry, etc \cite{doi:10.1021/acsnano.0c04702,schackert2015local,PhysRevB.99.064512,zhang2017origin,doi:10.1021/acsphotonics.1c00078,PhysRevB.95.024304,PhysRevB.99.165139,PhysRevB.74.184519,doi:10.1073/pnas.1406721111,PhysRevLett.130.256002}. In semiconductor quantum dots and wells, Fröhlich interaction between the charge and the electric field from the optical phonons is naturally enhanced when the chemical bonds are polar in nature \cite{wang2016tailoring}, but placing an interface, for example, an antiphase boundary, was shown to increase the Huang-Rhys factor (a measure of the EPC in optically excited semiconductors) by orders of magnitude even in nominally weakly polar III-V crystals \cite{doi:10.1021/acsnano.0c04702}. In metals, however, attempts to increase EPC by confining electrons \cite{schackert2015local}, confining phonons \cite{PhysRevB.99.064512}, interfacial charge transfer \cite{zhang2017origin}, enhanced electron surface scattering \cite{doi:10.1021/acsphotonics.1c00078,PhysRevLett.90.177401}, optical driving \cite{PhysRevB.95.024304}, or application of stress \cite{PhysRevB.99.165139,PhysRevB.74.184519,doi:10.1073/pnas.1406721111,PhysRevLett.130.256002} resulted only in moderate increase in $\lambda$ within a factor of $\sim$ two. An alternate strategy involves core@shell (e.g., Au@Ag or Au@Ag@Pt) nanostructures, where the effective EPC can be continuously tuned with core/shell mass fraction owing to sound velocity mismatch at the hetero-interface \cite{yu2015engineering}. Charge scattering at such heterointerfaces seems to determine the residual resistivity at low temperatures ($T$) in Ag@Au core@shell nanostructures~\cite{doi:10.1021/acsaelm.3c00379}, but their effect on the EPC has not been investigated. 

In this work, we have investigated EPC in noble metal hybrids consisting of a network of nanometer-sized Ag cores embedded in a crystalline Au matrix. The small intrinsic EPC and near-identical lattice constants of Ag and Au that preserve a global translation symmetry, provide a simple platform for analyzing the metallic state resistivity. Using electrical transport and point contact spectroscopy, we find that both static disorder and the EPC increase dramatically with increasing density of Ag nanoparticles (AgNP, core), {\it i.e.}, the proliferation of buried Ag-Au interfaces. At intermediate volume fractions of Ag, $\lambda$ as high as $\sim20$ could be observed, over ten times that of any known metal. This regime is also associated with a strong saturation in electrical resistivity that could be monitored by varying the EPC over nearly two orders of magnitude for the first time. 

Fig.~1a-d show the high-resolution transmission electron microscopy (HRTEM, Fig.~1a-b; TEM Fig.~1c) and scanning electron microscopy (SEM, Fig.~1d) image of the hybrid at increasing length scales. The building block consists of solution-processed $\sim 20 – 30$ nm Au shells each of which encloses multiple AgNPs of $\sim 2$ to $5$ nm diameter (Fig.~1a,b). The shells are subsequently fused, or ‘cross-linked’, and compacted to form macroscopic films on a glass substrate with pre-patterned electrical leads (Fig.~1e, Extended data Fig. 1). The Methods and Supplementary Information section I-III describe the chemical synthesis, characterization, and the film making processes in detail.  Fig.~1a (and also Extended data Fig. 2) emphasizes the sharp interface between AgNPs and the Au shell, which was found to be the case irrespective of the density $(\approx F/r_{\mathrm{Ag}}^3)$ of the AgNPs (here, $r_{\mathrm{Ag}}$ and $F= V_{\mathrm{Ag}}/(V_{\mathrm{Ag}}+V_{\mathrm{Au}})$ are the radius of the AgNP and the net relative volume fraction of Ag in the hybrid, respectively). Typically, we synthesize AgNPs of $r_{\mathrm{Ag}}\approx1-2$ nm and vary $F$ to tune the concentration of AgNPs, and thereby the inter-AgNP distance, $d_{\mathrm{Ag}}= 2r_{\mathrm{Ag}}/F^{1/3}$, and overall interface density, $F/r_{\mathrm{Ag}}$.

\begin{figure*}
\centering
  \includegraphics[height=7.5cm]{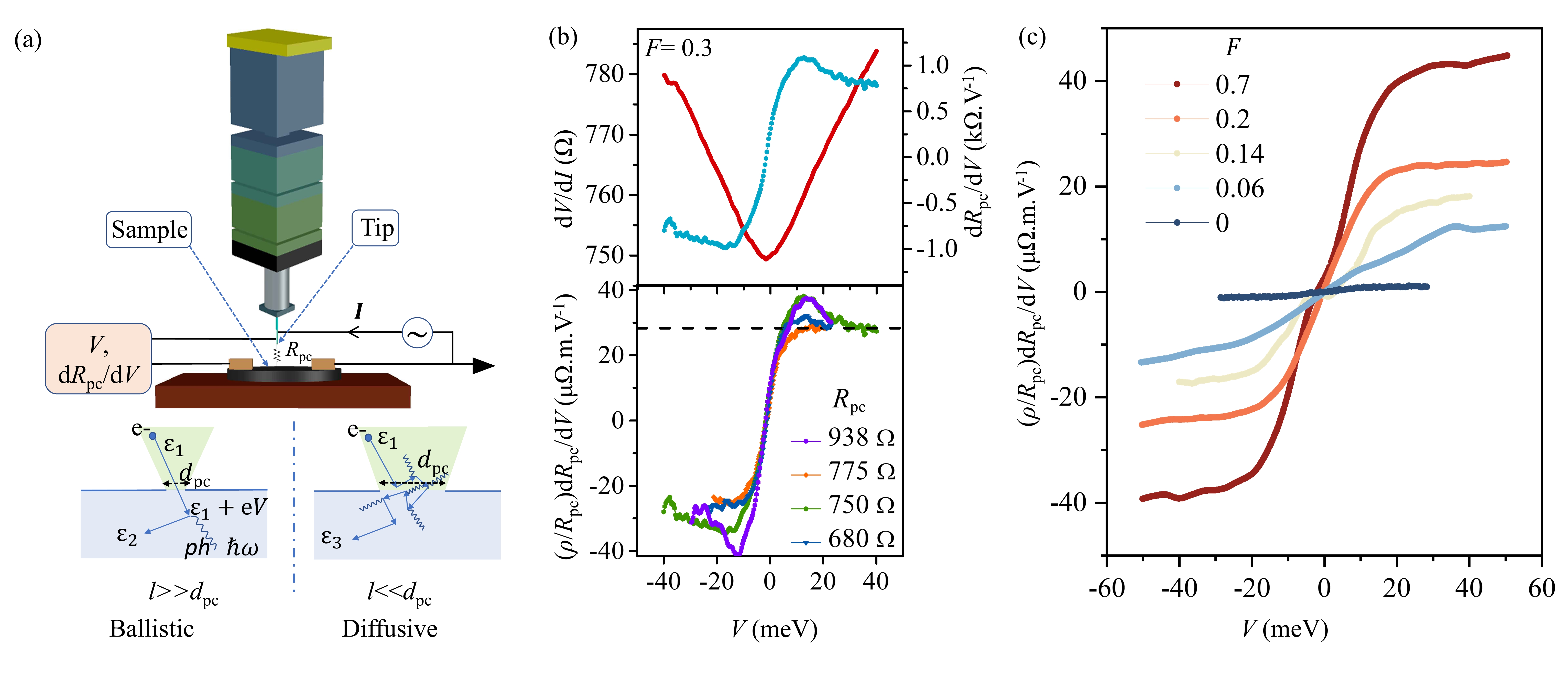}
 \caption{\textbf{Point contact spectroscopy of Ag@Au hybrid films:} \textcolor{black}{(a) Schematic and processes: The sensitivity of point contact spectroscopy to the phonon spectrum arises from the inelastic scattering of high energy electrons from an emitter (metallic tip) by the phonons in the sample. The tip is biased with a voltage ($V$) while the sample is grounded thereby driving current ($I$) across the tip-sample nanocontact of dimension $d_{\mathrm{pc}}$ and resistance $R_{\mathrm{pc}}$.
 An electron with initial energy $\epsilon_1$ gains an additional energy $eV$ while passing from the tip to the sample for a ballistic contact where $l\gg d_{\mathrm{pc}}$, $l$ being the mean free path of the electrons. Straight lines represent electrons, while curly lines indicate the emission of phonons responsible for the dissipation of the excess energy. In the case of a diffusive contact, since $l\ll d_{\mathrm{pc}}$ the energy relaxation at the orifice occurs via emitting multiple phonons, whose contributions accumulate towards a finite $\mathrm{d}R_{\mathrm{pc}}/\mathrm{d}V$ even at energies where the phonon density-of-states vanish ({\it i.e.} $eV \gg k_{\mathrm{B}}\Theta_{\mathrm{D}}$). The resulting weakly varying background at large $V$ is directly proportional to the electron-phonon coupling $\lambda$.}
 (b) Top panel shows the bias dependence of measured $\mathrm{d}V/\mathrm{d}I$ ({\it i.e.}, $R_{\mathrm{pc}}$) and $\mathrm{d}R_{\mathrm{pc}}/\mathrm{d}V$ for a typical film with Ag volume fraction $F=0.3$. The bottom panel shows the bias dependence of normalized point contact spectrum \textit{i.e.} $(\rho/R_{\mathrm{pc}})\mathrm{d}R_{\mathrm{pc}}/\mathrm{d}V$ for different point contact resistances which converge to a geometry-independent background (black dotted line). (c) Bias dependence of $(\rho/R_{\mathrm{pc}})\mathrm{d}R_{\mathrm{pc}}/\mathrm{d}V$ for films with different $F$ measured at $T\approx5$~K demonstrating an increasing background value with increasing $F$.}
  \label{fig2}
\end{figure*}
Fig.~1f shows the $T$-dependence of the electrical resistivity ($\rho$) for Ag@Au nanohybrid films of different $F$ between $6$~K and $300$~K. All the films showed metallic behavior where $\rho$ decreases monotonically with decreasing $T$ with very little or no evidence of upturn even at $T\sim0.3$~K (Extended data Fig.~3). Thorough compaction and crosslinking result in geometric uniformity and electrical homogeneity better than $\sim20\%$ (Extended data Fig.~1) and low background resistivity $\rho \sim 0.2~\mu\Omega.\mathrm{m}$ obtained in identically prepared films of bare Au nanoparticle (\textit{i.e.}, $F=0$). The incorporation of AgNPs causes $\rho$ to increase rapidly, which decreases again when the composite becomes Ag-rich ($F\rightarrow1$). This is seen in the variation in the residual resistivity $\rho_0$ (defined as $\rho$ at $T \approx 6$~K) with $F$, shown in Fig.~1g. Remarkably, in the intermediate range of $F\sim0.4 – 0.8$, $\rho_0$ is nearly constant and pinned to the magnitude of $\sim30-40~\mu\Omega.$m, which is scale of the Mott-Ioffe-Regel limit \cite{ioffe1960non}, $\rho_{\mathrm{MIR}}=3\pi^2 \hbar a/e^2\approx 10~\mu\Omega.m$ of metallic resistance for Au ($a = 0.3$~nm, is the lattice constant). At low $F$ ($\le 0.4$), $\rho_0$ increases linearly with the overall Ag-Au interface per unit volume, suggesting that the scattering of the electrons occurs dominantly at the buried Ag-Au interfaces (Extended data Fig.~4) \cite{doi:10.1021/acsaelm.3c00379}. 
\begin{figure*}
\centering
  \includegraphics[height=5.4cm]{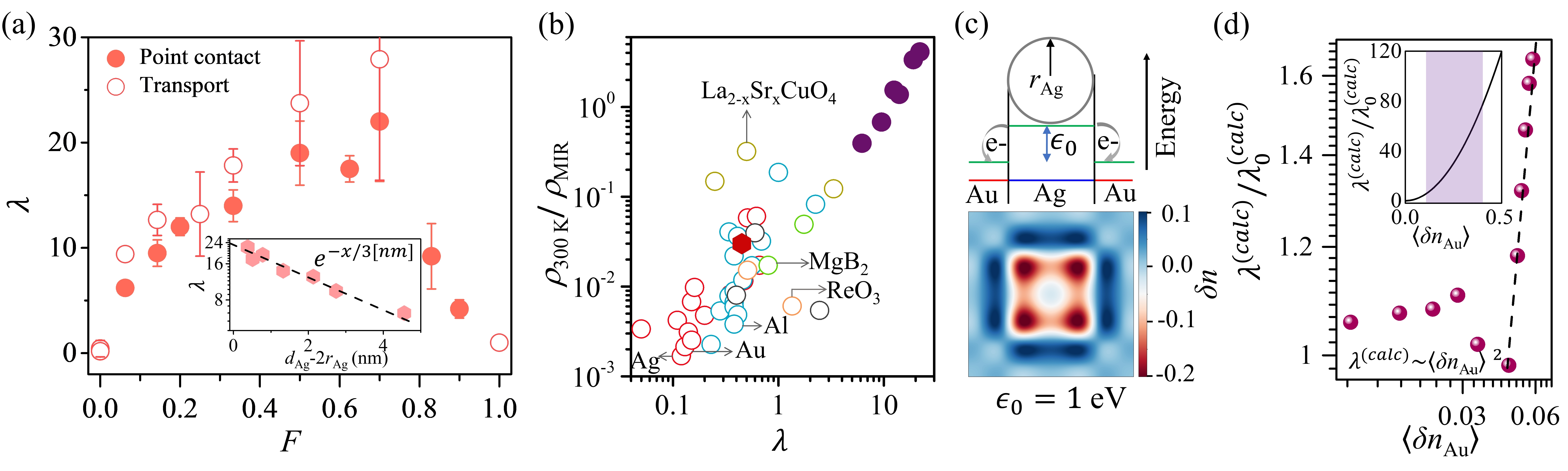}
 \caption{\textbf{Electron-phonon coupling in Ag@Au hybrids:} (a) Electron-phonon coupling constant ($\lambda$) estimated from point contact measurements and electrical transport ($\rho-T$) data as function of Ag filling $F$. Inset shows exponential decay of $\lambda$ with the distance between two AgNP-Au interfaces $\sim (d_{\mathrm{Ag}}-2r_{\mathrm{Ag}})$, where $d_{\mathrm{Ag}}$ and $r_{\mathrm{Ag}}$ are the average center-to-center distance and radius, respectively, of the AgNPs. The dashed line represents decay length scale $\sim3.2$~nm.(b) Room temperature resistivity ($\rho_{300~\mathrm{K}})$ for different materials is normalized by the respective Mott-Ioffe-Regel resistivity ($\rho_{\mathrm{MIR}}$) and plotted as a function of electron-phonon coupling constant $\lambda$. Red, blue, green, and yellow open circles represent non-superconducting metals, metals/alloys that superconduct at low $T$, intermetallic compounds, and high-$T_c$ cuprates, respectively (see Supplementary Information VI for detail). Filled red and purple circles represent films of pure Au nanoparticle and Ag@Au hybrids for different $F$ values,  respectively.
 \textcolor{black}{(c) (Top Panel):  A schematic of the electrochemical potential of electrons at the Ag and Au sites across the Ag@Au nanohybrid, $\epsilon_0$ being the potential difference between them. Electrons transfer from a higher onsite potential at Ag to a lower potential in Au. 
 (Bottom Panel): Theoretical computation of the excess electron occupancy $\delta n$ in a square lattice toy model (See Methods and Supplementary Information Section VII), where Ag is embedded inside Au. (d) Variation of the theoretically computed EPC, $\lambda^{\mathrm{(calc)}}$ with the average excess electron occupancy on Au, $\langle{\delta n_{\mathrm{Au}}}\rangle$. The dashed line shows a quadratic dependence of $\lambda^{\mathrm{(calc)}}$ with $\langle{\delta n_{\mathrm{Au}}}\rangle$. Inset shows the extrapolation of this quadratic dependence of $\lambda^{\mathrm{(calc)}}$ to higher values of $\langle{\delta n_{\mathrm{Au}}}\rangle\sim 0.5$. The shaded region of the inset depicts the regime of charge transfer found experimentally in the Ag@Au nanohybrids.}}
  \label{fig3}
\end{figure*}

The $T$-dependent component of electrical resistivity, {\it i.e.}, $\rho-\rho_0$, separately shown in Fig.~1h, contains two important features. First, the increase in the overall magnitude of $\rho-\rho_0$ with increasing $F$ at any given $T$, implies increasing contribution from phonons to resistivity, and thus enhancement in the ‘effective’ EPC. Secondly, incorporation of the AgNPs also makes the $T$-dependence of $\rho$ increasingly sublinear at high temperatures for $T>\Theta_\mathrm{D}$, where $\Theta_\mathrm{D}\sim 150$~K is the Debye temperature of Au. The coexistence of sub-linear $T$-dependence of $\rho$ in disordered metals with large EPC, such as the A15 compounds, has been known for many years \cite{PhysRevLett.38.782,gunnarsson2003colloquium}, but it is not expected in crystalline noble metals such as Au or Ag (or their alloys). The sublinearity makes the Bloch-Grüneisen formula ($\rho_{\mathrm{BG}} (T)$) for metallic resistivity, which derives $\rho\propto T$ for $T>\Theta_D$ by treating the EPC perturbatively, evidently inadequate except for $F = 0$ and $1$ (Extended data Fig.~5). This necessitates an alternative experimental tool to quantify the EPC parameter $\lambda$ in this case.

To estimate $\lambda$ independently, we have performed point contact spectroscopy on the Ag@Au nanohybrid films. Fig.~2a and 2b schematically explain the experimental arrangement and the underlying physical processes, respectively (see Methods and Extended data Fig.~6 for more detail). 
$\lambda$ is estimated from the energy-derivative of the resistance $R_{\mathrm{pc}}$ of a nanoscale contact between the film and a metallic Pt/Rh tip as (see derivation in Methods)~\cite{naidyuk2005point},

\begin{equation}
\lambda \approx \frac{3.4\pi ne\hbar}{16m} d_{\mathrm{pc}} \left[\frac{\mathrm{d}R_{\mathrm{pc}}}{\mathrm{d}V}\right]_{V\rightarrow\infty}.
\label{lambda}
\end{equation}

\noindent where $V$, $n$ and $m$ are the tip-sample bias, electron density in Au and electronic mass, respectively, and  $d_{\mathrm{pc}}=\rho/R_{\mathrm{pc}}$ is the contact diameter (Maxwell regime). The upper panel of Fig.~2c shows typical $V$-dependence of $R_{{\mathrm{pc}}}$ and $\mathrm{d}R_{\mathrm{pc}}/\mathrm{d}V$. At large $V$, {\it i.e.} $e|V|\gg\epsilon_t$, where $\epsilon_\mathrm{t} \sim k_\mathrm{B} \Theta_\mathrm{D} \sim 10-20$ meV is the energy scale beyond which the Migdal-Eliashberg function (Eq.~\ref{alpha_calc}) $\alpha^2 \mathcal{F}(\omega)\rightarrow 0$, $d_{\mathrm{pc}}\mathrm{d}R_{\mathrm{pc}}/\mathrm{d}V$ becomes independent of the geometric details of the contact, as seen from the convergence of the traces at different $R_{\mathrm{pc}}$ (lower panel of Fig.~2c). Fig.~2d shows the $V$-dependence of  $d_{\mathrm{pc}}\mathrm{d}R_{{\mathrm{pc}}}/{d}V$ for different values of $F$. The rapid increase in its large-$V$ magnitude with increasing $F$ confirms the enhancement in the EPC with the incorporation of AgNPs. 

Fig.~3a shows the magnitude of $\lambda$, obtained from Eq.~\ref{lambda} using $d_{\mathrm{pc}}\mathrm{d}R_{\mathrm{pc}}/\mathrm{d}V$ values at large $V$, as function of $F$. Remarkably, we find $\lambda$ can be as high as $\approx 20$ for $F\sim 0.5 – 0.7$, before dropping to $\sim 1$, expected for a film of small AgNPs \cite{PhysRevLett.90.177401}. Both the magnitude and the $F$-dependence of $\lambda$ from point contact spectroscopy are consistent within $20\%$ with the estimates obtained by fitting the Bloch-Grüneisen formula at low $T$ ($\le 100$~K) (open symbols in Fig.~3a, see Methods and Extended data Fig.~5 for details). Such large $\lambda$ is unprecedented in metallic solids, and exceeds those with strong EPC, for example the A15 compounds, by at least a factor of $\sim$ ten (Fig.~3b, also see Supplementary Information section VI for details). Intriguingly, the extended correlation between $\lambda$ and the normalized resistivity in Fig.~3b suggests Ag@Au hybrids to be an `extreme' case of a metal where the electron-phonon scattering drives $\rho \rightarrow \rho_{\mathrm{MIR}}$ even at room temperature.

\textcolor{black}{Surface scattering can increase EPC in nanostructured metal films compared to bulk~\cite{doi:10.1021/acsphotonics.1c00078}, but such enhancements are within a factor of $\sim$ two, much smaller than the enhancement observed here. The near-exponential decay of $\lambda$ with the separation, $d_{\mathrm{Ag}}-2r_{\mathrm{Ag}}$, between neighbouring Ag-Au interfaces (inset of Fig.~3a), however, suggests the effect could arise from a (screened) Coulomb interaction between the interfaces and the conduction electrons. Indeed, the formation of the solution processed Ag@Au core-shell nanoparticle hybrids, and their stability against galvanic replacement, are critically dependent on the interfacial charge transfer that results in to electric dipoles across the hetero-interface~\cite{MOTT201214,doi:10.1021/j100173a053,10.1063/1.3626031,doi:10.1021/acs.chemmater.1c04176}. The X-ray photoelectron spectroscopy (XPS) in these nanohybrids suggests average electron doping of the Au atoms by as much as $\sim 0.3-0.4$ per atom for $F = 0.5$ (Extended Data Fig.~7 and Supplementary Information II.B). To verify if interfacial charge transfer can indeed cause substantial enhancement in $\lambda$, we computed the EPC using a two-dimensional lattice with a periodic square array of Ag atoms embedded within a matrix of Au atoms where the on-site charge excess charge ($\delta n$) depends on the difference ($\epsilon_0$) between the Ag and Au site energies and the long-range Coulomb interaction between electrons (Fig.~3d (schematic), Methods and Supplementary Information Section VII). The radial dipoles across the interface, formed when the Ag$^{|\delta n|+}$ and Au$^{|\delta n|-}$ sites (Fig.~3d, lower panel) assume opposite oxidation states, couple strongly to the lattice phonons via long-range Coulomb interactions and generate additional contributions to the Migdal-Eliashberg function:}

\begin{equation}
\label{alpha_calc}
    \alpha^2\mathcal{F}(\omega) = A\sum_{k,q}|g(k,q)|^2\delta(\varepsilon_{k})\delta(\varepsilon_{k+q})\delta(\omega - \omega_q)
\end{equation}

\noindent
\textcolor{black}{where $k$ is the electron wave vector, $q$ and $\omega_q$ are the phonon wave vector and frequency, respectively; $A$ is a normalization constant, and $g(k,q)$ are the the electron-phonon matrix elements. The nanostructuring and the resulting charge redistribution generates additional contribution to $g \propto (\delta n)V_0$, where $V_0$ embodies the strength of the inter-site Coulomb interaction and $\delta n$ is the on-site excess electron occupancy (see Methods and Supplementary Information Section VII for the analytical and computational details). Fig.~3e shows the calculated EPC, $\lambda^{\mathrm{(calc)}}$, normalized to its value for $\delta n \rightarrow 0$, as a function of the average electron transfer $\langle\delta n_{Au}\rangle$ to the Au atoms. We note that $\lambda^{\mathrm{(calc)}}$ increases nearly by a factor of $\sim$ two within the calculated range $\langle\delta n_{Au}\rangle \leq 0.07$ (limited by the finite lattice size), and second, $\lambda^{\mathrm{(calc)}} \propto \langle\delta n_{Au}\rangle^2$ at larger $\langle\delta n_{Au}\rangle$. The latter, when extrapolated to the experimentally estimated $\langle\delta n_{Au}\rangle \sim 0.4$, suggests an enhancement of $\lambda^{\mathrm{(calc)}}$ (shaded region in the inset of Fig.~3e), which is within a factor of $\sim2-3$ of that observed in experiments. }

We now focus on the resistivity saturation at $T\gg\Theta_\mathrm{D}$, which probably is the ‘smoking gun’ signature of the strong emergent EPC in Ag@Au hybrids. In fact, our ability to vary $\lambda$ by over a factor $\sim 200$ (from bare gold film to Ag@Au nanohybrid at $F\approx0.7$), allows access to the phonon contribution to resistivity dynamically from weak to ultra-strong coupling regime on a single material platform for the first time. In Fig.~4, we plotted the high-temperature segment ($300$~K $\ge T \ge 150$~K, i.e., $T\ge \Theta_\mathrm{D}$) of ($\rho-\rho_0$) shown in Fig.~1h for all $F$, where the temperature axis is scaled by the corresponding $\lambda$, obtained from the point contact measurements. Two key observations can be summarized as follows: First, the collapse of the resistivity traces for different $F$ onto a single one suggests $\lambda T$ would continue to be the ‘scaling variable’ that determines the resistivity even at very large $\lambda$, although the perturbative limit with linear scattering rate $\approx2\pi k_\mathrm{B} \lambda T/\hbar$, is expectedly recovered only when $\lambda\rightarrow 0$ (dashed line). Second, the sublinearity in $(\rho-\rho_0)$ at large $\lambda T$, representing ‘resistivity saturation’, can be modeled with a parallel resistor channel as,

\begin{equation}
    \frac{1}{\rho-\rho_0}=\frac{1}{\rho_{\mathrm{BG}}}+\frac{1}{\rho_{||}}
    \label{fit_RT}
\end{equation}

\noindent where $\rho_{\mathrm{BG}} (T\ge \Theta_\mathrm{D})=2\pi m k_\mathrm{B} \lambda T/ne^2\hbar$ \cite{PBAllen,ziman1972principles} (See Methods for details), and $\rho_{||}$ is the resistivity of a parallel non-classical channel whose universality, temperature dependence, or even existence, have been questioned many times, but without a satisfactory answer so far \cite{PhysRevLett.38.782,gunnarsson2003colloquium,werman2016mott}. The solid line fit in Fig.~4 corresponds to $\rho_{||}\approx 20~\mu\Omega.$m implying that the phonon contribution to resistivity can be described by an `ideal’ Bloch-Grüneisen behavior in parallel to a $T$-independent non-classical channel of resistivity close to the MIR-limit. In fact, the variation in $\rho(T)$ over the entire experimental temperature range can be satisfactorily captured by using the full form of $\rho_{\mathrm{BG}} (T)$ and Eq.~\ref{fit_RT} (dashed lines in Fig.~1h). See Methods and Supplementary Information Section V for further discussions on the parallel resistor formula and other fit protocols.

\begin{figure}
\centering
  \includegraphics[height=5.3cm]{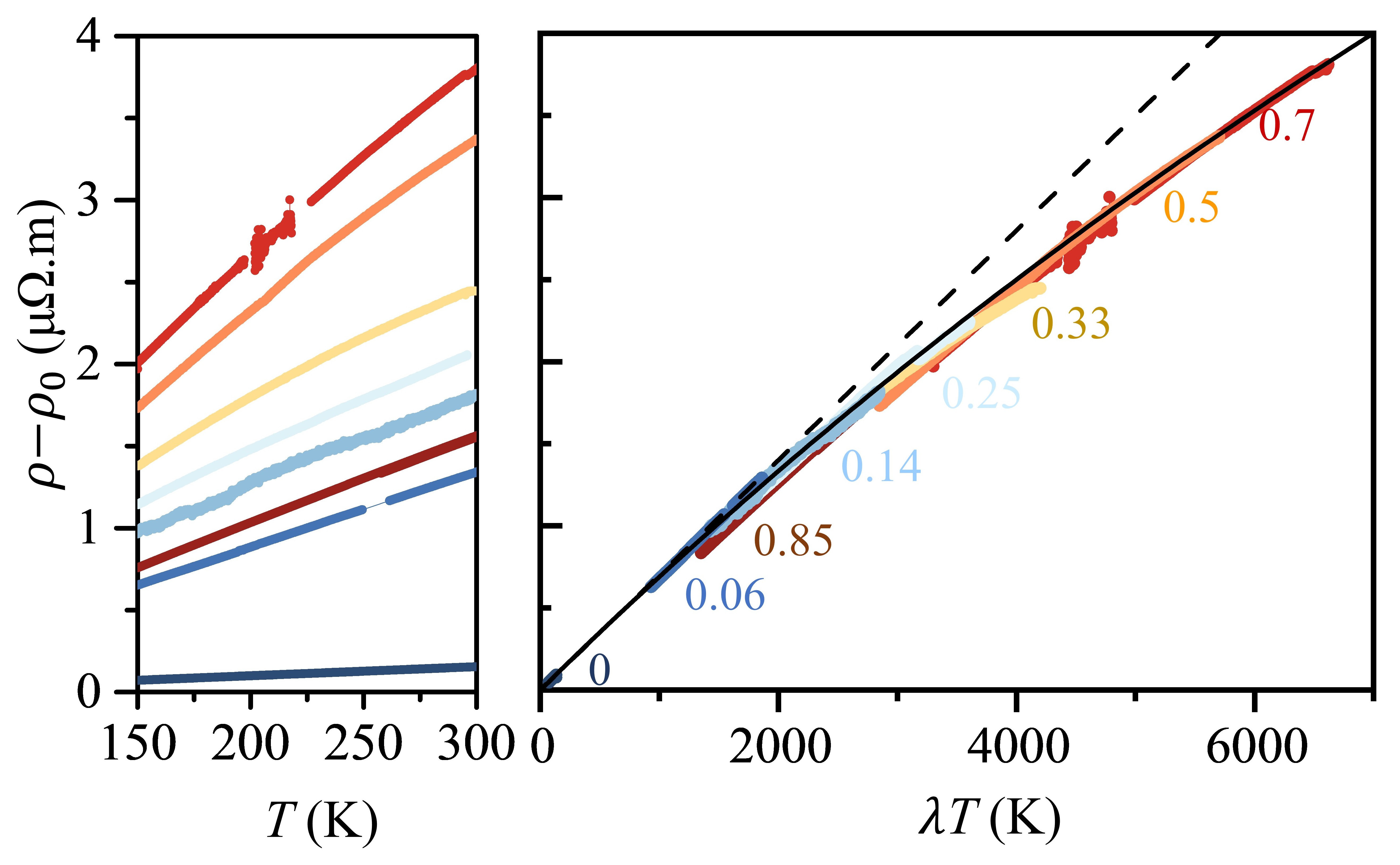}
 \caption{\textbf{Scaling of resisitivity and electron-phonon coupling strength:} Universality in the $T$-dependent component of $\rho$ for $T>\Theta_\mathrm{D}\sim150$~K (left panel) for all $F$ with $T$ scaled with the corresponding EPC parameter $\lambda$ from point contact spectroscopy (right panel). The dashed line in the right panel indicates the `Planckian' resistivity $(= 2m\pi k_\mathrm{B}\lambda T/(\hbar ne^2))$. The solid line is a fit to $(\rho - \rho_0)$ according to the two-component parallel channel model given by Eq.~[\ref{fit_RT}] (see text).
 }
\label{fig4}
\end{figure}

Our experiment confirms the long-suspected inevitability of resistivity saturation in metals \cite{ALLEN1980291}, irrespective of the strength of the EPC. However, there are deeper consequences. The first concerns the question, is there a universal bound to the EPC for a metal to exist? The persistence of metallic transport in Ag@Au hybrids with $\lambda\gg 1$, is a priori at odds with a stability bound observed in Monte Carlo calculations on the Holstein model \cite{doi:10.1073/pnas.2216241120}, or other fundamental ‘Planckian limits’ to dissipation or thermalization in metals \cite{doi:10.1126/science.1227612,PhysRevLett.123.066601}. An important consideration would, however, be the heterogeneous nature of our system, where the electrons couple to engineered vibration modes of a foreign species, rather than those of the host lattice itself. \textcolor{black}{While in-homogenous polaronic instabilities could lead to a thermally activated parallel channel in transport \cite{werman2016mott,werman2017non} (See Supplementary Information Section IV), the apparent suppression of the quantum correction to conductivity, and indeed that of Anderson localization itself, even for very strong disorder at $F \sim 0.5 – 0.8$ at low  $T$ (down to $\sim0.3$ K) are unexpected.} These suggest a significant modification of the phase coherent effects in our nanohybrids, whose origin is not clear at present but inelastic scattering at the buried interfaces could have a role to play. Third, there is also a discrepancy with the models of resistivity saturation built on the breakdown of Born-Oppenheimer approximation and Matthiessen’s rule, for example, polaronic deformation of disorder \cite{ciuchi2018origin} or \textcolor{black}{phonon-assisted delocalization} \cite{PhysRevB.68.033103,werman2016mott,werman2017non}. These mechanisms often involve crossover to negative temperature coefficient of resistivity at $\rho\sim\rho_{\mathrm{MIR}}$, i.e., `Mooij correlation' \cite{mooij1973electrical,ciuchi2018origin}, which was not observed in any of our samples. Finally, we doubt that electron-electron interaction, which can impact transport in multiple ways including assisting delocalization of carriers \cite{PhysRevLett.73.2607}, introducing hydrodynamic viscosity \cite{PhysRevLett.106.256804} or suppression of resistivity saturation itself \cite{gunnarsson2003colloquium,PhysRevLett.38.782,hussey2004universality}, is of importance in our case. This is because the charge carrier density in the hybrids, measured by the low-field Hall effect (Extended data Fig. 8), is found to be the same as that of crystalline Au in which electronic interaction is known to be weak.

In conclusion, we have reported the realization of a metallic hybrid, composed of ultra-small silver cores dispersed in a crystalline gold matrix, in which the electrical resistivity shows saturation as the silver core density and temperature are increased. Both electrical transport and point contact spectroscopy reveal that the electron-phonon coupling $\lambda$ in these engineered metallic hybrids can be as large as $\sim 20$, more than ten times than any known metallic solid. 
\textcolor{black}{Our experiments outline a novel strategy to modify some of the fundamental properties of metallic solids utilizing buried interfaces at the nanoscale.}

{$^\star$\textbf{Equal contribution}}\\
 {$^\dagger$\textbf{Corresponding authors}:\\
shreyak@iisc.ac.in\\
tuhinmaji@iisc.ac.in\\
arindam@iisc.ac.in}
\bibliography{bibliography}

\newpage
\section*{Methods}
\subsection*{Chemical synthesis}
Ag@Au nanohybrids (Ag@Au NHs) were synthesized using a colloidal approach~\cite{doi:10.1021/acsaelm.3c00379,Saha_2022}. The synthesis process involved two sequential stages: the reduction of AgNO$_3$ with ice-cold NaBH$_4$ to form Ag nanoparticles (AgNPs) in an aqueous solution (in ultrapure water, purity $\sim18.2~\mathrm{M}\Omega$.cm) containing NaOH, NH$_4$Br, KI, and CTAB as a capping agent, followed by the introduction of HAuCl$_4$ at $40\hspace{0.05cm}^\circ$C with continuous stirring. In situ, UV-Vis absorption spectroscopy was employed to monitor the synthesis process, and the reaction was terminated by adding isopropyl alcohol (IPA), resulting in nanohybrid agglomeration. The solution was then centrifuged at $10,000$ rpm for $15$ minutes to remove excess CTAB and isolate the nanohybrids.

\subsection*{Characterization}
\noindent \textbf{UV-Vis Spectroscopy:} UV-vis absorption spectroscopy was used to assess the formation of AgNPs and Ag@Ag nanohybrids. The addition of NaBH$_4$ in the reaction mixture containing AgNO$_3$, at a specific 'wait time' ($t$) resulted in a prominent peak around $\sim393$ nm, corresponding to the localized surface plasmon resonance (LSPR) absorption band of ultra-small AgNPs. Upon introducing HAuCl$_4$, the LSPR peak underwent a redshift to $\sim524$ nm over approximately $1000$ seconds, indicating the formation of a thicker shell due to spontaneous interdiffusion. To ensure a well-defined AgNP surface, we employed a strategic approach. Monitoring the SPR band shift with UV-Vis spectroscopy, we terminated the reaction by adding isopropyl alcohol (IPA) approximately $30$ seconds after HAuCl$_4$ addition.

\noindent \textbf{X-Ray Photo-electron Spectroscopy:} An Axis Ultra K$\alpha$ X-ray photoelectron spectrometer with a monochromatized photon energy of $\sim 1486.6$ eV was used for all XPS measurements. To minimize moisture absorption, the sample was quickly inserted into the load-lock of the instrument, pumped in the entry chamber until the pressure around $10^{–8}$ mbar was reached, and subsequently transferred to the analysis chamber. The individual core-level spectra were corrected for charging using C $1s$ at $284.5$ eV as standard. The peak fitting of the individual core-levels was done using Casa XPS software. 

\noindent\textbf{Transmission Electron Microscopy:} Transmission Electron Microscopy (TEM) was employed to examine the structural characteristics of an Ag@Au nanohybrid. The FEI TITAN Themis TEM operating at $300$ kV, which offers a point resolution of $\sim 0.2$ nm and an energy spread of $0.136$ nm, was utilized to capture all TEM images. High-Resolution Transmission Electron Microscopy (HRTEM) images were obtained along the zone axis to assess the Ag-Au interface and investigate any defects that may have arisen within the bimetallic entities. To prepare the sample for TEM imaging, the nanoparticle underwent multiple cleaning cycles in a CHCl$_3$ : methanol mixture ($1:3$ ratio), followed by centrifugation at $15000$ rpm for $5$ minutes. The resulting precipitated sample was then dispersed in chloroform and deposited onto a carbon-coated TEM grid, which was subsequently dried under vacuum overnight.

\subsection*{Film preparation}
The drop-cast technique was employed to fabricate the Ag@Au NH film onto prepatterned Cr$/$Au contacts (with thickness of $\approx$ 10 nm$/$60 nm) arranged in various lead configurations on a glass substrate. Prior to drop-casting onto the pre-patterned leads, the sample was dissolved in CHCl$_3$. Subsequently, the sample was dried at 70 $^\circ$C for 30 seconds and washed with deionized (DI) water followed by KOH solution and IPA to eliminate any excess CTAB and achieve a chemically sintered cross-linked nanostructure. This process was repeated ten times for each film, resulting in an average film thickness of $t_\mathrm{f} \approx 3 \pm 0.5~ \mu$m  and a diameter of  $\approx4$ mm, typically covering the leads (Fig.~1e of Main Manuscript, Extended Data Fig.~1). 


\subsection*{Electrical Measurement}
Four-probe resistivity of the sample was measured down to temperature ($T$) $\sim6$~K in a home-built cryostat by passing a DC current of $\sim100~\mu$A with Keithley 6221 and measuring the voltage with Keithley 2182A. Voltage across multiple contacts were recorded using Keithley 3700 Multiplexer card. The voltage was measured in delta mode to cancel any thermo-emf across the contacts. Resistivity from $T\sim10$~K down to $T\sim0.3$~K was measured in a He3 cryostat.
\subsection*{Fitting of $\rho-T$ data}
The resistivity ($\rho$) of metal with electron-phonon interaction playing the dominant role of scattering can be expressed in terms of the Bloch Grüneisen \cite{PBAllen,ziman1972principles} form as:
\begin{equation}
    \rho(T)=\rho_0+\rho_{\mathrm{BG}}(T)
    \label{BG}
\end{equation}
where $\rho_0$ is the residual resistivity, and 
\begin{equation}
\rho_{\mathrm{BG}}=\frac{2\pi\lambda k_\mathrm{B}/\Theta_{\mathrm{D}}}{(n/m) e^2}\left(\frac{T}{\Theta_\mathrm{D}}\right)^5\int_0^{\Theta_\mathrm{D}/T}\frac {x^5}{(e^x-1)(1-e^{-x})}dx
\label{rho_BG}
\end{equation}
is the Bloch Grüneisen form of resistivity arising from electron-phonon scattering. $\Theta_\mathrm{D}$, the Debye temperature, and $\lambda$, the electron-phonon coupling constant can be estimated by fitting the $\rho-T$ data with Eq.~[\ref{BG}].
As shown in Fig.~1(h) and Extended Data Figure 5(a), we have fitted the $\rho-T$ data of AuNP and AgNP films using Eq.~[\ref{BG}]. $\Theta_{\mathrm{D}}\sim170$~K, and $\lambda\sim0.45$ for Au, and $\Theta_{\mathrm{D}}\sim190$~K and $\lambda\sim1$ for Ag are estimated as fit parameters. The increased value of $\lambda$ as compared to the bulk value of $\sim0.2$ for Au and Ag could be attributed to the nanostructuring in the film and increased electron scattering from the surfaces \cite{doi:10.1021/acsphotonics.1c00078}. 
For Ag@Au films, Eq.~[\ref{BG}] cannot describe $\rho-T$ for the entire range of $T$. Extended data Fig.~5(b) shows that the transport data for $F=0.5$ deviates from the low-temperature BG fit ($T\le100~\mathrm{K})$ to the data.
For fitting the $\rho-T$ data for Ag@Au hybrid films, Eq.~[2] in main manuscript is used where $\rho_{\parallel}$, $\Theta_\mathrm{D}$ and $\lambda$ are the parameters of fit. $\Theta_\mathrm{D}$ obtained from fitting $\rho-T$ of Ag@Au films with Eq.~[\ref{BG}] in the low $T$ range, and with Eq.~[2] with a parallel conduction channel, are consistent and lies within the range of $150-170$~K for all films (Extended Data Figure 5(c)). $\Theta_D$ being close to the Debye temperature of Au ($\Theta_{\mathrm{D,Au}}\sim170$~K) in all cases indicates the electrical conduction occurs primarily within the host lattice of Au. $\lambda$ derived from the low-$T$ Bloch Grüneisen fit is shown in Fig.~3(a). However, $\lambda$ estimated from parallel channel fit is slightly overestimated ($\sim20\%$) and probably less accurate since the scattering mechanism with this model even at low $T$ is not purely electron-phonon mediated. 
$\rho_{\parallel}$ estimated from the parallel channel fit, given by Eq.~[2] of main manuscript, is plotted in Extended Data Figure 5(d) as a function of $F$ after normalizing with $\rho_{\mathrm{MIR}}$. $\rho_{\parallel}\approx 8-25~\mu\Omega.$m is $T$-independent, and lies within a factor of two of $\rho_{\mathrm{MIR}}\sim10~\mu\Omega$.m, which 
drives the saturation of $\rho(T)$.

\subsection*{Point contact Measurements}
\noindent {\bf Experimental setup:} A sharp Pt/Rh metallic tip is brought in contact with the film in a controlled manner with the help of nanopositioners (attocubes and piezo tubes) as indicated in the schematic of the experimental set-up in Fig.~2(a) of main manuscript and Extended Data Figure.~6. The tip-sample chamber is loaded inside a home-built cryostat that could be cooled down to $T\sim5$~K. 

\noindent {\bf Theoretical framework for analysis:} Point contact measurements are done with the technique of modulation spectroscopy \cite{naidyuk2005point,10.1116/6.0001087} which measures the higher-order derivatives of a signal through its AC components at harmonics of a definite frequency.
The circuit for measurement of the point contact voltage is shown in Extended data Fig.~6.
A mixed AC+DC current $I+i_\mathrm{m} \mathrm{sin}(\omega t)$ is passed through the sample using a constant current circuit. AC voltage from a lock-in amplifier and DC voltage from Keithley 2400 are added with an op-amp adder. A series resistance $R_\mathrm{s}\gg R_{\mathrm{pc}}$ is used to achieve the constant current in the circuit.
The voltage across the tip-sample contact can be represented as a Taylor series expansion of the $I-V$ curve as follows:
\begin{align}
    V={}&f(I+i_\mathrm{m} \mathrm{sin}(\omega t))=\sum_{k=1}^{\infty}\frac{i^{k}_\mathrm{m}}{k!}\frac{\mathrm{d}V^k}{\mathrm{d}^kI}\mathrm{sin}^{k} (\omega t)\nonumber\\
    ={}&V+i_\mathrm{m} \mathrm{sin}(\omega t)\frac{\mathrm{d}V}{\mathrm{d}I}+\frac{1}{2!}\frac{\mathrm{d}^2V}{\mathrm{d}I^2}i_\mathrm{m}^2 \mathrm{sin}^2 (\omega t)\nonumber\\
    {}&+\frac{1}{3!}\frac{\mathrm{d}^3V}{\mathrm{d}I^3}i_\mathrm{m}^3 \mathrm{sin}^3 (\omega t)+...\nonumber\\
    ={}&1\left(V+\frac{i_\mathrm{m}^2}{4}\frac{\mathrm{d}^2V}{\mathrm{d}I^2}+...\right)+\nonumber\\
    {}&\mathrm{sin}(\omega t)\left(i_\mathrm{m}\frac{\mathrm{d}V}{\mathrm{d}I}+\frac{i_\mathrm{m}^3}{8}\frac{\mathrm{d}^3V}{\mathrm{d}I^3}+...\right)-\nonumber\\
    {}&\mathrm{cos}(2\omega t)\left(\frac{i_\mathrm{m}^2}{4}\frac{\mathrm{d}^2V}{\mathrm{d}I^2}+\frac{i_\mathrm{m}^4}{12}\frac{\mathrm{d}^4V}{\mathrm{d}I^4}+...\right)+..
    \label{taylor_exp}
\end{align}

Grouping terms of the same amplitude of the modulation frequency $\omega$ gives
\begin{equation}
    V=V_0+\sum_{i=n}^{\infty}(a_{2n-1}\mathrm{sin}((2n-1)\omega t)+a_{2n}\mathrm{cos}(2n\omega t))
\end{equation}
For a small AC current $i_\mathrm{m}\ll I$, the higher-order terms in the series expansion, which vary as $i^n$ can be neglected in each group and the amplitude of the voltage at frequency $n\omega$ becomes proportional to the $n$-th order derivative \textit{.i.e} $a_n\propto {\mathrm{d}^nV}/{\mathrm{d}I^n}$. Also, it is to be noted that the odd harmonics are in-phase (sine component) with the source AC signal $i_\mathrm{m}\mathrm{sin}(\omega t)$, and even harmonics are out of phase, at $90^\circ$ (cosine component) with the signal.\\
Hence, the first order derivative can be estimated from the amplitude of the $\omega$ component at $0^\circ$ phase as in Eq.~[\ref{taylor_exp}].
\begin{align}
    a_1={}&i_\mathrm{m}\frac{\mathrm{d}V}{\mathrm{d}I}\nonumber\\
    \frac{\mathrm{d}V}{\mathrm{d}I}={}&\frac{a_1}{i_\mathrm{m}}
    \label{dV/dI}
\end{align}
The second order derivative can be derived from the amplitude of the $2\omega$ component at $90^\circ$ phase as in Eq.~[\ref{taylor_exp}].
\begin{align}
    a_2={}&-\frac {i_\mathrm{m}^2}{4}\frac{\mathrm{d}^2V}{\mathrm{d}I^2}\nonumber\\
    \frac{\mathrm{d}^2V}{\mathrm{d}I^2}={}&-\frac{4a_2}{i_\mathrm{m}^2}
    \label{d2V_dI2}
\end{align}
Both $\omega$ and $2\omega$ components of the voltage difference across the tip-sample junction are acquired simultaneously with two lock-in amplifiers as denoted by $V_\omega$, and $V_{2\omega}$ in Extended Data Fig.~6 respectively, after amplification with SR 560. The sample is grounded through current-voltage amplifier SR 570, which allows us to constantly monitor the current through the sample and hence tune the point contact resistance ($R_{\mathrm{pc}}$).\\ 
Since a lock-in amplifier shows the rms value of a signal, the amplitude of the signal is $\sqrt{2}$ times the measured value \textit{i.e.} $a_n=V_{n\omega}\sqrt{2}$, where $V_{n\omega}$ is the measured signal at $n\omega$. If $V_{\mathrm{AC}}$ is the voltage at the sine output of the lockin amplifier, the AC current through the sample is $i_{\mathrm{m}}=V_{\mathrm{AC}}\sqrt{2}$. $V_{\omega}$, being the measured voltage by the lock-in amplifier at $\omega$ component and $0^{\circ}$ phase, the amplitude of the voltage drop across the tip and sample is expressed using Eq.~[\ref{dV/dI}] as
\begin{equation}
\frac{\mathrm{d}V}{\mathrm{d}I}= \frac{a_1}{i_\mathrm{m}}=\frac{V_{\omega}\sqrt{2}}{V_{\mathrm{AC}}\sqrt{2}/R_\mathrm{s}}=\frac{V_{\omega}}{V_{\mathrm{AC}}/R_\mathrm{s}}
\label{dV_dI_final}
\end{equation}
Similarly, $V_{2\omega}$, being the signal measured by the lockin amplifier at $2\omega$ component and $90^{\circ}$ phase, the second order derivative is expressed using Eq.~[\ref{d2V_dI2}] as 
\begin{equation}
\frac{\mathrm{d}^2V}{\mathrm{d}I^2}=-\frac{4V_{2\omega}\sqrt{2}}{(\sqrt{2} V_{\mathrm{AC}}/R_\mathrm{s})^2}=-\frac{4V_{2\omega}}{\sqrt{2}(V_{\mathrm{AC}}/R_\mathrm{s})^2}
\label{d2V_dI2_final}
\end{equation}
$\mathrm{d}V/\mathrm{d}I$ corresponds to the point contact resistance $R_{\mathrm{pc}}$ and $\mathrm{d}^2V/\mathrm{d}I^2$ is related to the derivative of $R_{\mathrm{pc}}$ with bias as $\mathrm{d^2}V/\mathrm{d}I^2=R_{\mathrm{pc}} \mathrm{d}R_{\mathrm{pc}}/\mathrm{d}V$. 

With appropriate multi-stage vibration isolation, $R_{\mathrm{pc}}$ ranging from 500 $\Omega$
to $2$~K$\Omega$ could be stabilized typically in the Ag@Au hybrid films with attocube and piezo controllers by monitoring the current through the sample. The modulation AC current $i_\mathrm{m}$ was typically fixed at $1-5~\mu$A, whereas the DC current $I$ was varied till $\sim 100-200~\mu$A in magnitude. $R_{\mathrm{pc}}$ can be expressed as a combination of ballistic Sharvin resistance ($R_{\mathrm{sh}}={16\rho l}/{3\pi d_\mathrm{pc}^2}$) and diffusive Maxwell resistance \cite{naidyuk2005point} ($R_{\mathrm{M}}=\rho/d_\mathrm{pc}$).
\begin{equation}
    R_{\mathrm{pc}}=\frac{16\rho l}{3\pi d_{\mathrm{pc}}^2}+\frac{\rho}{d_\mathrm{pc}}
    \label{Rpc}
\end{equation}
where $v_\mathrm{F}$ is the Fermi velocity, and $d_{\mathrm{pc}}$ is the diameter of the point contact orifice as shown in the schematic of Fig.~2(b) of the main manuscript.
Using a typical resistivity of $10~\mu\Omega$.m and Drude mean free path $l\approx0.1$~nm, we get $d_{\mathrm{pc}}\approx10$~nm for $R_{\mathrm{pc}}\approx~1$~k$\Omega$ from Eq.~[\ref{Rpc}]. This indicates a diffusive nature of the contact since $l\ll d_{\mathrm{pc}}$. For an ideal ballistic point contact, the derivative of the point contact resistance represents the Migdal Eliashberg spectral function $g(\epsilon)=\alpha^2\mathcal{F}(\epsilon)$ \cite{naidyuk2005point}, which is the probability of specific phonon modes (with energy $\epsilon$) to decay into an electron-hole pair and resembles closely with the phonon density of states in most cases. 
\begin{equation}
    \frac{1}{R_{\mathrm{sh}}}\frac{\mathrm{d}R_{\mathrm{pc}}}{\mathrm{d}V}=\frac{8ed_{\mathrm{pc}}}{3\hbar v_\mathrm{F}}g(\epsilon)|_{\epsilon=eV}
    \label{ballistic_PC}
\end{equation}
The integral of $g(\epsilon)$ is a measure of the electron-phonon coupling constant. 
\begin{equation}
\lambda=2\int^\infty_0\frac{g(\epsilon)}{\epsilon} d\epsilon
\label{lambda}
\end{equation}
Hence, from the measurement of the point contact spectrum, we can estimate the electron-phonon coupling parameter.

In the diffusive regime, the spectrum is modified as \cite{naidyuk2005point} 
\begin{subequations}
\begin{align}
\frac{1}{R_{\mathrm{sh}}}\frac{\mathrm{d}R_{\mathrm{pc}}}{\mathrm{d}V}=&\frac{8ed_{\mathrm{pc}}}{3\hbar v_F}[g(\epsilon)+
\gamma\int_0^{\infty} \frac{g(\omega)}{\omega+\omega_{0}}d\omega+\\
&\frac{\gamma}{2} \frac{eV}{eV+\omega_0}g(\epsilon)]_{\epsilon=eV}
\end{align}
\end{subequations}
$\gamma\approx0.58$ is a geometrical factor arising from the shape of the orifice. The last two terms represent correction to the ballistic expression due to a background signal arising from the successive generation of phonons by the energized electrons due to increased collisions at the orifice ($d_{\mathrm{pc}}\gg l$) and then a further scattering of these phonons from electrons or defects at the point contact. For strong reabsorption of phonons, \cite{naidyuk2005point}, the phonon escape frequency $\omega_0=\omega_\mathrm{D} l_{\mathrm{ph}}l_{\mathrm{r}}/d_{\mathrm{pc}}^2$ is an order or less than the phonon Debye frequency ($\omega_\mathrm{D}$) as $d_{\mathrm{pc}}\gg l_{\mathrm{ph}},l_{\mathrm{r}}$, with $l_{\mathrm{ph}}$, and $l_\mathrm{r}$ being the inelastic phonon-electron scattering length and elastic phonon-defect scattering lengths respectively. In this case, $\omega_0\rightarrow0$. Since, $g(e|V|\gg\hbar\omega_\mathrm{D})=0$, the point contact spectrum can be approximated as follows: 
\begin{subequations}
\begin{align}
    \frac{1}{R_{\mathrm{sh}}}\frac{\mathrm{d}R_{\mathrm{pc}}}{\mathrm{d}V}&\approx\frac{8ed_{\mathrm{pc}}}{3\hbar v_\mathrm{F}}g(\epsilon),  \\
    &  \epsilon \le \hbar\omega_\mathrm{D}, \text{Ballistic regime}\nonumber\\
    &\approx\frac{8ed_{\mathrm{pc}}}{3\hbar v_\mathrm{F}}[g(\epsilon)[1+\frac{\gamma}{2}]+\gamma\frac{\lambda}{2}], \nonumber\\
    &  \epsilon \le \hbar\omega_\mathrm{D}, \text{Diffusive regime}\\
    &\approx0,\\
    &\epsilon\ge\hbar\omega_\mathrm{D}, \text{Ballistic regime}\nonumber\\
    &\approx\frac{8ed_{\mathrm{pc}}}{3\hbar v_\mathrm{F}}\frac{\gamma}{2}\lambda, \\
    & \epsilon\ge\hbar\omega_\mathrm{D}, \text{Diffusive regime}\nonumber
\end{align}
\label{diffusive_PC}
\end{subequations}
The magnitude of the point contact spectrum depends on the contact dimension $d_{\mathrm{pc}}$. Hence after normalizing the experimentally measured value of $\mathrm{d}R_{\mathrm{pc}}/\mathrm{d}V$ with $d_{\mathrm{pc}}$, we get a quantity independent of the contact geometry and reflecting $g(\epsilon)$ or $\lambda$, the fundamental material properties.
Hence, we simplify the above equations as:
\begin{subequations}
\begin{align}
    {d_{\mathrm{pc}}}\frac{\mathrm{d}R_{\mathrm{pc}}}{\mathrm{d}V} & \approx \frac{16 m}{\pi n e \hbar}g(\epsilon), \nonumber \\
    &\hbar\omega_\mathrm{D}\le eV, \text {Ballistic regime}\\
    &\approx \frac{16 m}{\pi n e \hbar}[g(\epsilon)[1+\frac{\gamma}{2}]+\gamma\frac{\lambda}{2}], \nonumber\\ 
    &\hbar\omega_\mathrm{D}\le eV, \text {Diffusive regime}\\
    &=0, \nonumber\\
    &  \hbar\omega_\mathrm{D}\gg eV, \text{Ballistic regime}\\
    &\approx \frac{16 m}{\pi n e \hbar}\frac{\lambda}{3.4},\nonumber\\
    &  \hbar\omega_\mathrm{D}\le eV, \text{Diffusive regime}
    \label{beta_dR_dV_2}
\end{align}
\end{subequations}
For a ballistic point contact, it is thus possible to study the phonon spectrum in $g(\epsilon)$.
However, in a diffusive regime for strong reabsorption of phonons, there is a non-zero background which makes it difficult to distinguish the phonon peaks in $g(\epsilon)$.
Nonetheless, $\lambda$ can be evaluated from the background value ($e|V|\gg\epsilon_\mathrm{t}\sim\hbar\omega_\mathrm{D}$) of $d_{\mathrm{pc}} \mathrm{d}R_{\mathrm{pc}}/\mathrm{d}V$ at $eV\gg\hbar\omega_\mathrm{D}$.
In the diffusive regime, the point contact resistance can be approximated by the Maxwell resistance \textit{i.e} $d_{\mathrm{pc}}\approx\rho/R_{\mathrm{pc}}$. Hence, for a given $R_{\mathrm{pc}}$ and $\rho$, where a small corrective factor $F^{-1/3}$ (usually $\lesssim 20$\% for most films), is multiplied to the latter for $F > 0$ to account for the difference in the  effective `surface' disorder as opposed to that in the bulk transport, we can estimate $d_{\mathrm{pc}}$ using the above relation and following Eq.~[\ref{beta_dR_dV_2}] derive the value of $\lambda$ from the normalized $\mathrm{d}R_{\mathrm{pc}}/\mathrm{d}V$ as:
\begin{equation}
    \lambda=3.4\frac{\pi n e\hbar}{16 m} \left[\frac{\rho}{R_{\mathrm{pc}}}\frac{\mathrm{d}R_{\mathrm{pc}}}{\mathrm{d}V}\right]_{V\rightarrow\infty}
\end{equation}
The diffusive nature of the point-contact is experimentally established by the scaling of $\mathrm{d}R_{\mathrm{pc}}/\mathrm{d}V$ at different values of $R_{\mathrm{pc}}$ (and hence $d_{\mathrm{pc}}$) assuming $d_{\mathrm{pc}}=\rho/R_{\mathrm{pc}}$. 
\section*{D\MakeLowercase{ata availability}}
Source data are provided with this paper. Data that support the plots within this paper, and other findings of this study are available from the corresponding author upon reasonable request. 
\section*{C\MakeLowercase{ode availability}}
The codes that support the findings of this study are available from the
corresponding author upon reasonable request.
\section*{C\MakeLowercase{ompeting interests}}
The authors declare no competing interests.
\setcounter{figure}{0}
\renewcommand{\figurename}{\textbf{Extended Data Fig.}}
\renewcommand{\thefigure}{\textbf{\arabic{figure}}}

\newpage

\begin{figure*}[t!]
  \includegraphics[height=16cm]{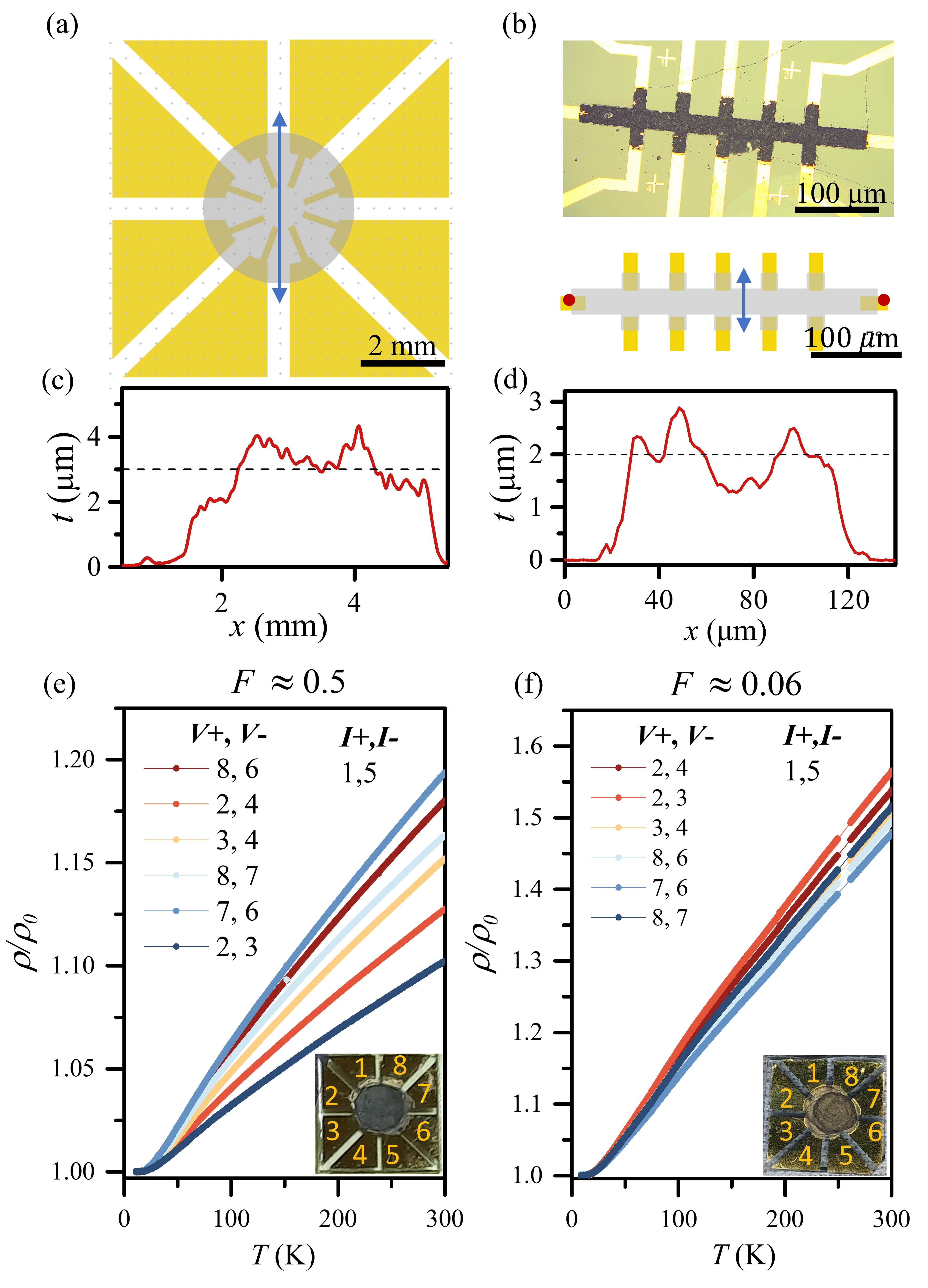}
 \caption{\textbf{Film Uniformity:} \textcolor{black}{(a). Schematic of Van der Pauw leads \cite{philips1958method} used for  most electrical measurements. The gold-coloured regions indicate the deposited Cr/Au electrodes and the grey circle indicates the dropcast film. (b) Optical image of a typical film on Hall bar leads. The bottom panel shows the schematic of the Hall bar leads, with red dots indicating the current contacts. The intermediate leads have been used for resistivity measurements as indicated by the green points of Fig.~1g of the main manuscript. All configurations of leads, provide identical results in resistivity measurements. (c) and (d) show the thickness profiles of typical films in Van der Pauw and Hall bar geometry measured via optical profilometry along the blue lines indicated in Fig.~1a and bottom panel of Fig.~1b, respectively. $x$ represents the spatial extent of the film. The average thickness is $2-3~\mu$m, represented by the dotted lines.} (d) and (e) show the temperature ($T$) dependence of the resistivity ($\rho$) normalized by the residual resistivity ($\rho_0$, defined as the resistivity at $T\sim 6$~K) along multiple voltage channels for Ag filling $F=V_{\mathrm{Ag}}/(V_{\mathrm{Ag}}+V_{\mathrm{Au}})=0.5$ and $0.06$, respectively. $I+, I-$, and $V+, V-$ indicate the current and voltage probes respectively. Insets show the typical images of the films at respective values of $F$. Variation of $\rho/\rho_0$ is within $10\%$ along different channels which was observed in all films.}
  \label{Extended Data Fig.1}
\end{figure*}
\newpage
\begin{figure*}[t!]
  \includegraphics[height=7cm]{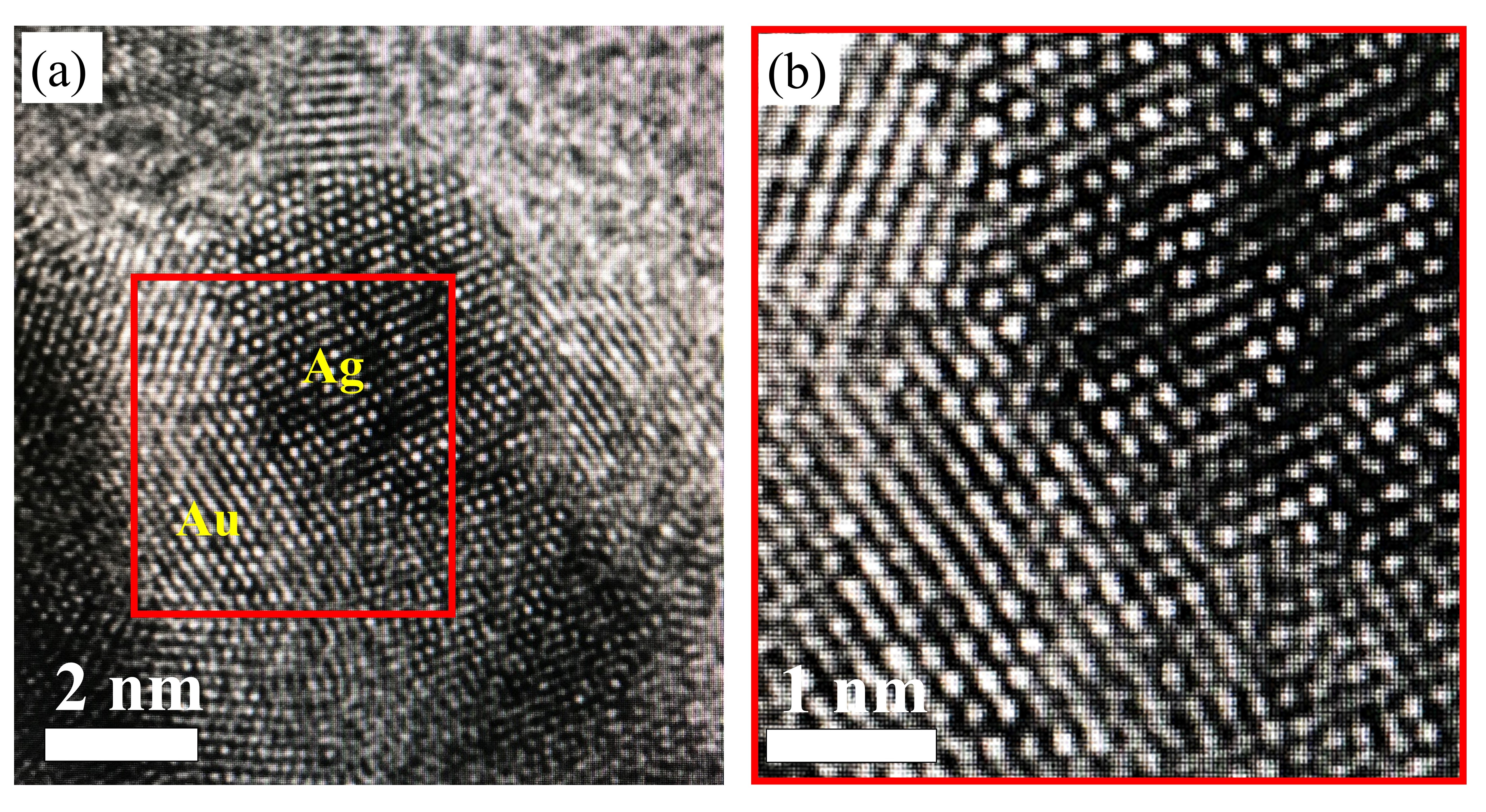}
 \caption{\textbf{High-Resolution Transmission Electron Microscopy (HRTEM) Image of Ag@Au Interface:} (a) presents a high-resolution image of the Ag@Au nanohybrid with atomic precision, where the darker region represents the AgNPs surrounded by the lighter region, the Au matrix. The zoomed-in image (b) focuses on the interface between the AgNPs and the Au matrix, marked by a red squre in (a), revealing a well-defined sharp boundary between the AgNPs and the Au matrix. This indicates that the AgNPs are encapsulated by the Au matrix, forming a core-shell structure. The absence of visual line defects in these images indicates the high-quality crystalline nature of the nanohybrid.}
  \label{Extended Data Fig.2}
\end{figure*}
\newpage
\begin{figure*}[t!]
\centering
  \includegraphics[height=16cm]{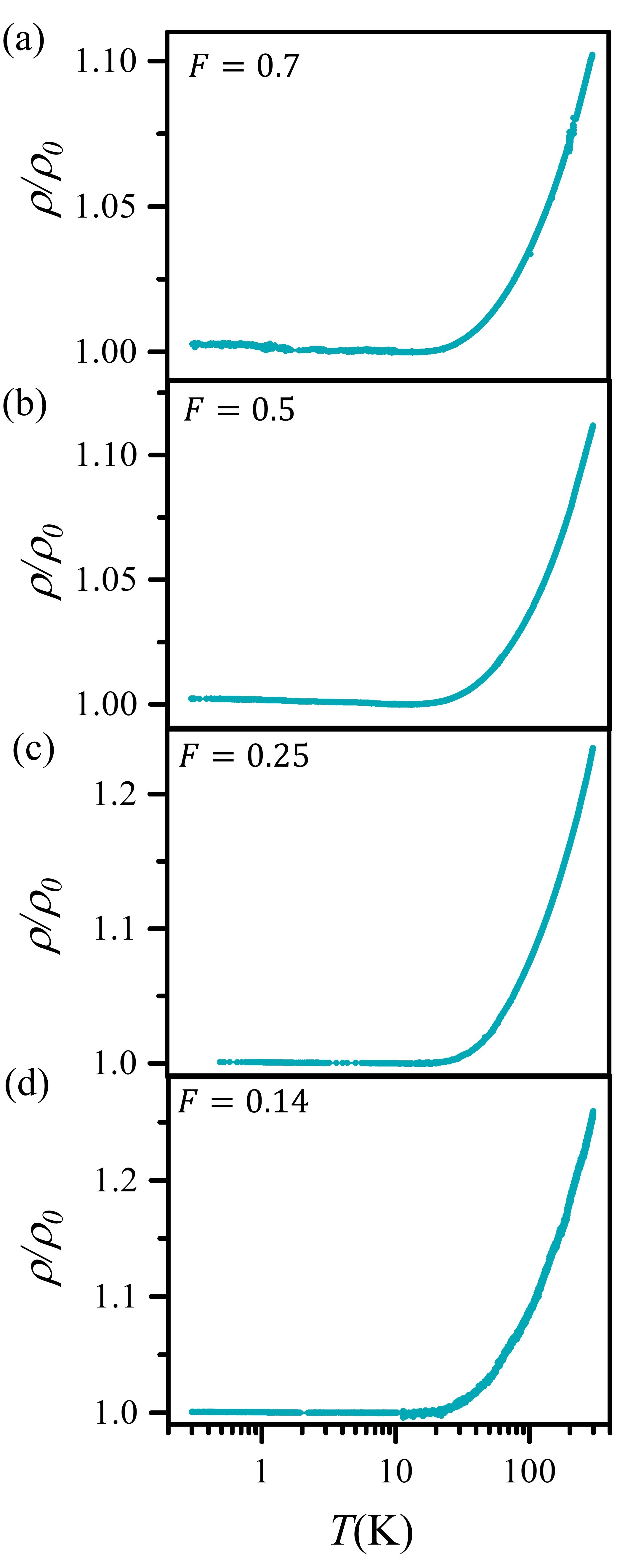}
 \caption{\textbf{Low-temperature transport data:} (a), (b), (c), and (d) represent the temperature ($T$) dependence of the average normalized resistivity ($\rho/\rho_0$) for films with different Ag filling $F=0.7, 0.5, 0.25$, and $0.14$, respectively. The resistivity measurements were performed down to $T\sim 0.3$~K. Irrespective of $F$, no evidence of an upturn in resistance with decreasing $T$ (negative temperature coefficient of resistance), e.g., due to activation of electrons across inter-grain tunnel barrier through tunneling or variable range hopping in case of weakly interconnected or granular assembly of nanoparticles \cite{PhysRevB.56.10596,PhysRevLett.107.176803,PhysRevB.89.041406,PhysRevB.76.212201,PhysRevB.78.075437,doi:10.1021/nn101376u,duan2013controllability,simon1998charge}, or any other disorder-mediated quantum coherent processes such as weak localization \cite{PhysRevB.80.245318}, was observed down to $T\sim0.3$~K. \textcolor{black}{The upper limit in the correction to conductivity ($\sigma$) for $F\sim0.5$ is $\sim100~\Omega^{-1}.\mathrm{m}^{-1}$, which is at least an order of magnitude smaller than that for bare Au film.
 }
 }
  \label{Extended Data Fig.3}
\end{figure*}
\newpage
\begin{figure*}[t!]
\centering
  \includegraphics[height=6cm]{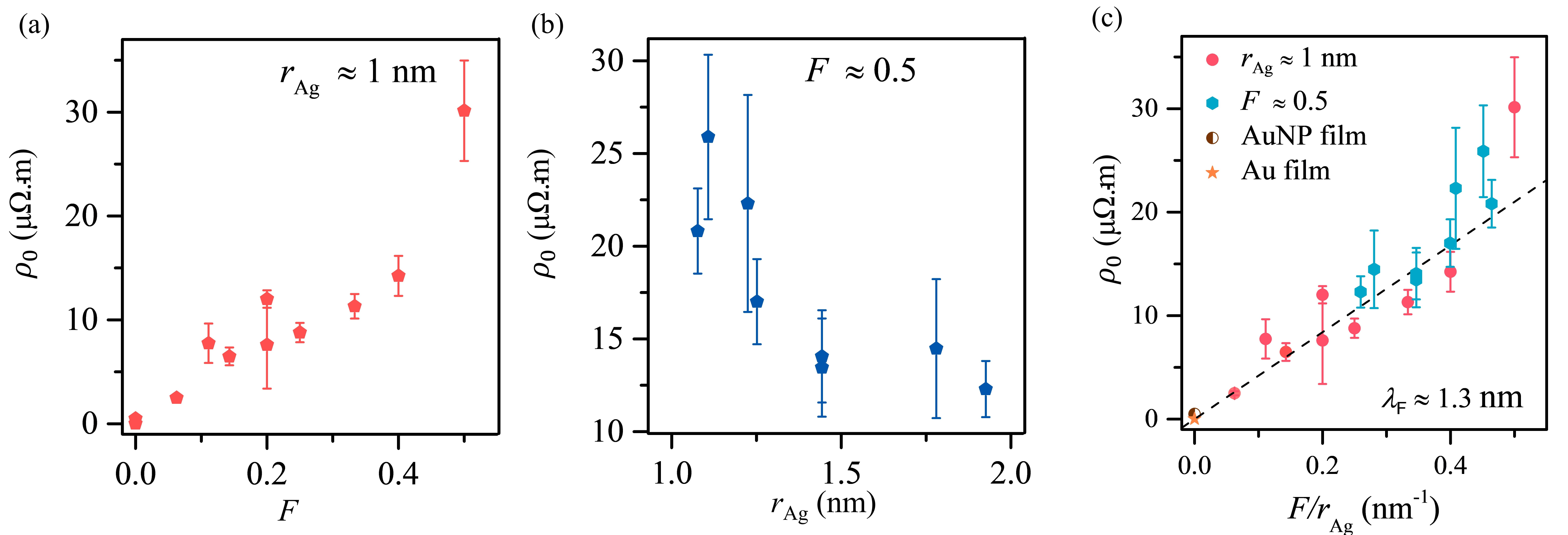}
 \caption{\textbf{Variation of residual resistivity with the density of Ag-Au interface area at low fractional filling ($F \lesssim 0.5$):} (a) Residual resistivity ($\rho_0$), defined as the resistivity at $T\sim6$~K, as a function of the volume fraction $F$ of Ag at the fixed AgNP radius ($r_{\mathrm{Ag}} \approx 1$~nm). (b) $\rho_0$ as a function of $r_{\mathrm{Ag}}$ at a fixed $F (\approx 0.5)$. (c) $\rho_0$  as a function of $F/r_{\mathrm{Ag}}$, the interface area of the AgNP per unit volume of the hybrid film. The collapse of all points onto a single linear behavior in $F/r_{\mathrm{Ag}}$ indicates that the Ag-Au interfaces are the dominant sources of scattering of electrons, as opposed to random atomic scale defects or alloying-related disorder. Within a simple Landauer-Büttiker formalism~\cite{doi:10.1021/acsaelm.3c00379}, where $\rho_0 \approx (h\lambda_\mathrm{F}^2/e^2)\times(F/r_\mathrm{Ag})$, we get the Fermi wavelength $\lambda_\mathrm{F}\sim1.3$~nm (dashed line), which is within a factor of $\sim 2 - 3$ of that of crystalline Au.}
  \label{Extended Data Fig.4}
\end{figure*}
\newpage
\begin{figure*}[t!]
\centering
  \includegraphics[height=12cm]{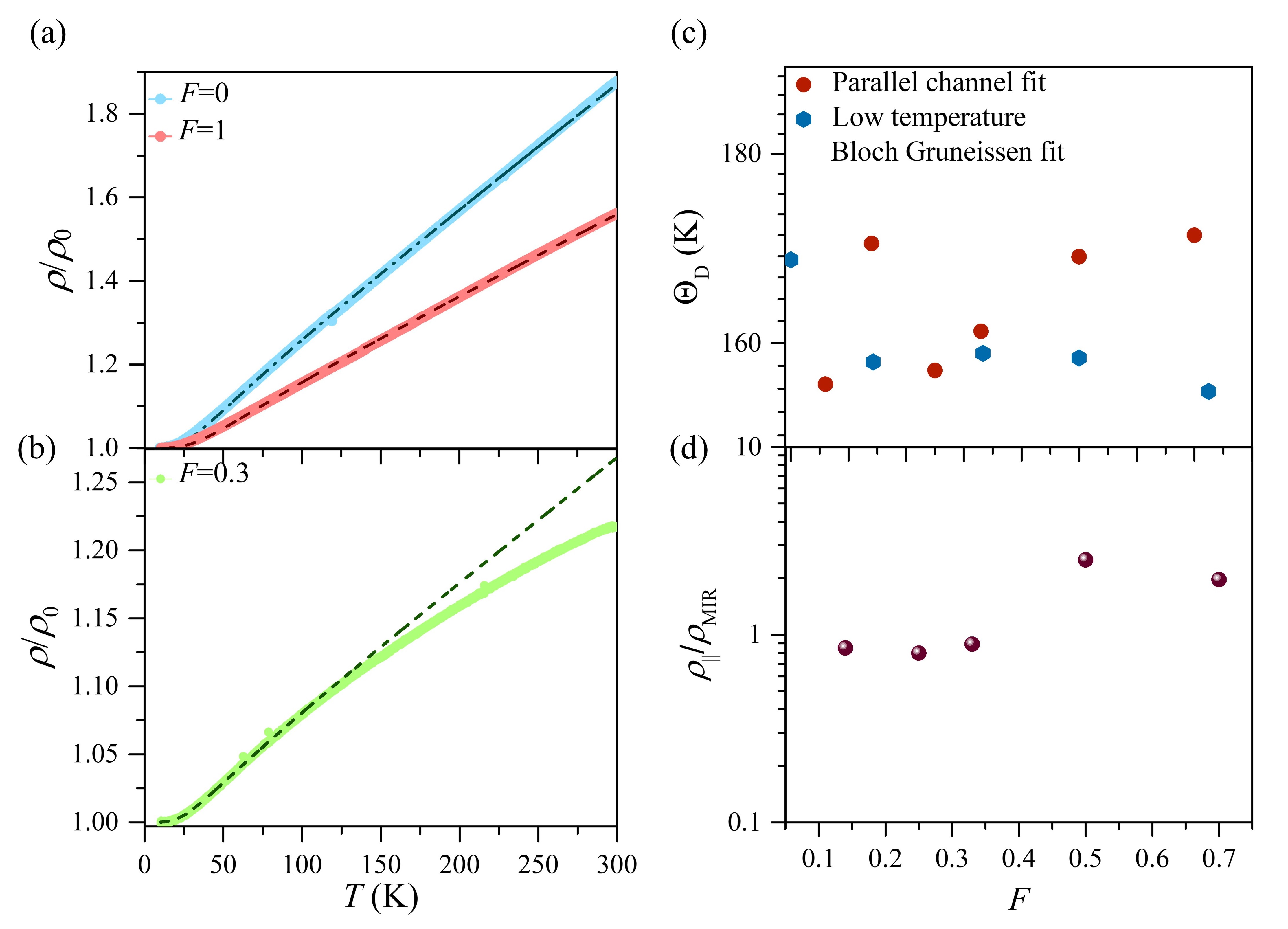}
 \caption{\textbf{Fit of $\rho-T$ data:} (a) Bloch-Grüneisen fit (using Eq.~[\ref{BG}]) to $\rho-T$ data for Au nanoparticle ($F=0$) and Ag nanoparticle ($F=1$) films. Electron-phonon coupling constant $\lambda \approx 0.45$, and Debye temperature
$\Theta_\mathrm{D}\approx 160~\mathrm{K}$ for Au, and $\lambda \approx 1$ and $\Theta_\mathrm{D}\approx 190~\mathrm{K}$ for Ag are derived as fit parameters. (b) Eq.~[\ref{BG}] is used to fit the $\rho-T$ data for $T\le 100$~K in a film of Ag fraction $F=V_{\mathrm{Ag}}/(V_{\mathrm{Ag}}+V_{\mathrm{Au}})=0.33$ giving $\Theta_\mathrm{D}\sim 160$~K and $\lambda\sim18$. However, extrapolation of the fit to high $T>100$~K shows the deviation of the data from linearity. 
The sub-linearity of $\rho–T$ for $T\geq \Theta_\mathrm{D}$ naturally indicates the inadequacy of the Bloch-Grüneisen fit for films with $0 < F < 1$. (c) $\Theta_\mathrm{D}$ obtained from the  Bloch-Grüneisen fit to $\rho-T$ data at $T<100$~K and parallel channel model for entire $T$ range is plotted as a function of $F$. $\Theta_\mathrm{D}$ varies between $150-170$~K in all cases, which is close to that of Au ($\Theta_{\mathrm{D,Au}}\sim180$~K). (d) $T$-independent parallel channel resistivity ($\rho_{||}$), obtained as a fit parameter in the parallel channel fit (Eq.~[2] of main manuscript) to $\rho-T$ data is plotted with $F$ after normalizing with Mott-Ioffe-Regel resistivity \cite{ioffe1960non} $\rho_{\mathrm{MIR}}\approx3\pi^2\hbar^2a/e^2\approx10~\mu\Omega$.m.}
  \label{Extended Data Fig.5}
\end{figure*}
\newpage
\begin{figure*}[t!]
\centering
  \includegraphics[height=7.4cm]{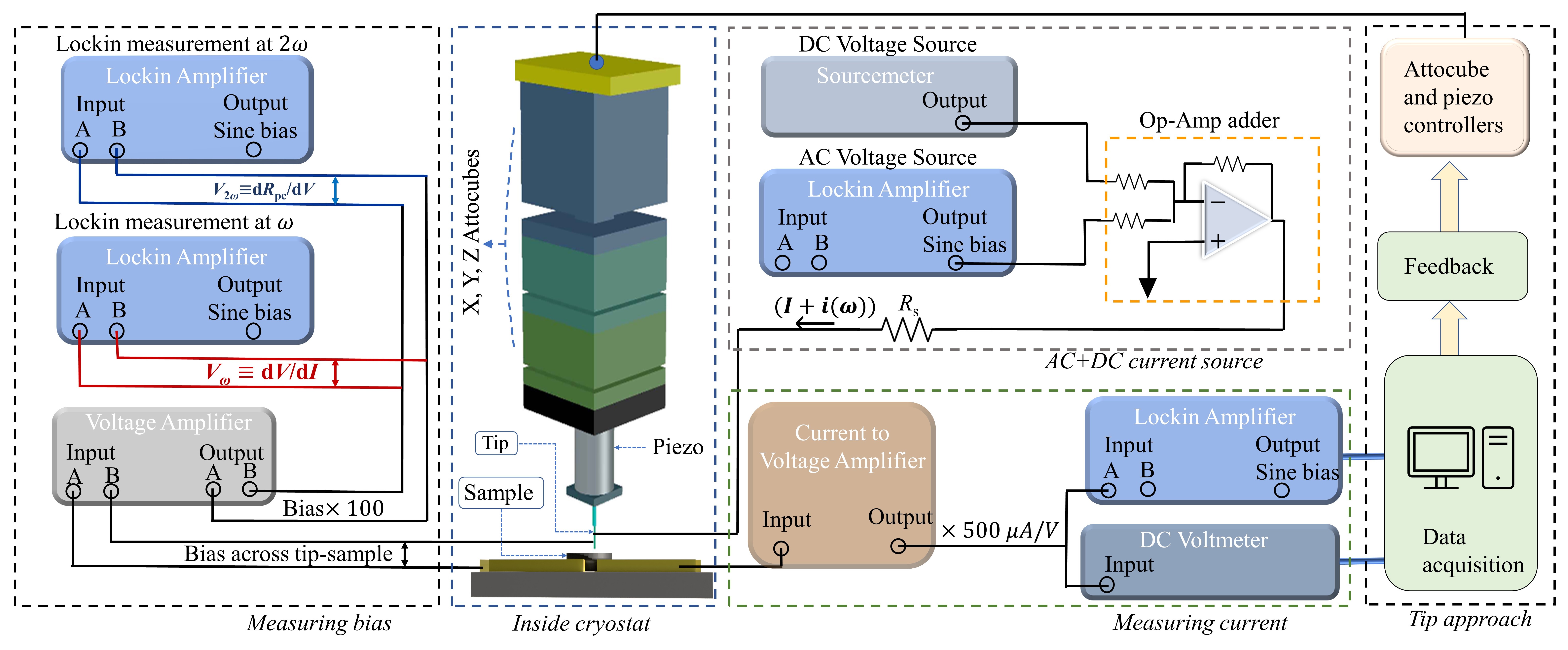}
 \caption{\textbf{Schematic of the electrical circuit for point-contact measurements:} The tip is mounted on an attachment comprising of attocubes and piezo, which can control the movement of the tip in x,y and z direction down to the precision of a nanometer. The sample prepared on pre-patterned glass substrate is loaded under the tip. This entire sample-tip chamber is loaded inside a home-built cryostat which can go down to temperature $\sim5$~K. Mixed AC+DC current ($I+i(\omega)$, $\omega$ being the frequency of the AC component) is passed to the tip by using a voltage source and series resistance $R_\mathrm{s}$. The voltage across the tip-sample contact is then amplified with a voltage amplifier and measured with two lockin amplifiers simultaneously at frequencies $\omega$ and $2\omega$. The sample is grounded through a current-to-voltage amplifier, which converts even a small amount of current through the sample into a measurable voltage drop. The current through the sample is measured to monitor the tip-sample resistance.}
  \label{Extended Data Fig.6}
\end{figure*}
\newpage
\begin{figure*}[t!]
\centering
  \includegraphics[height=12.5cm]{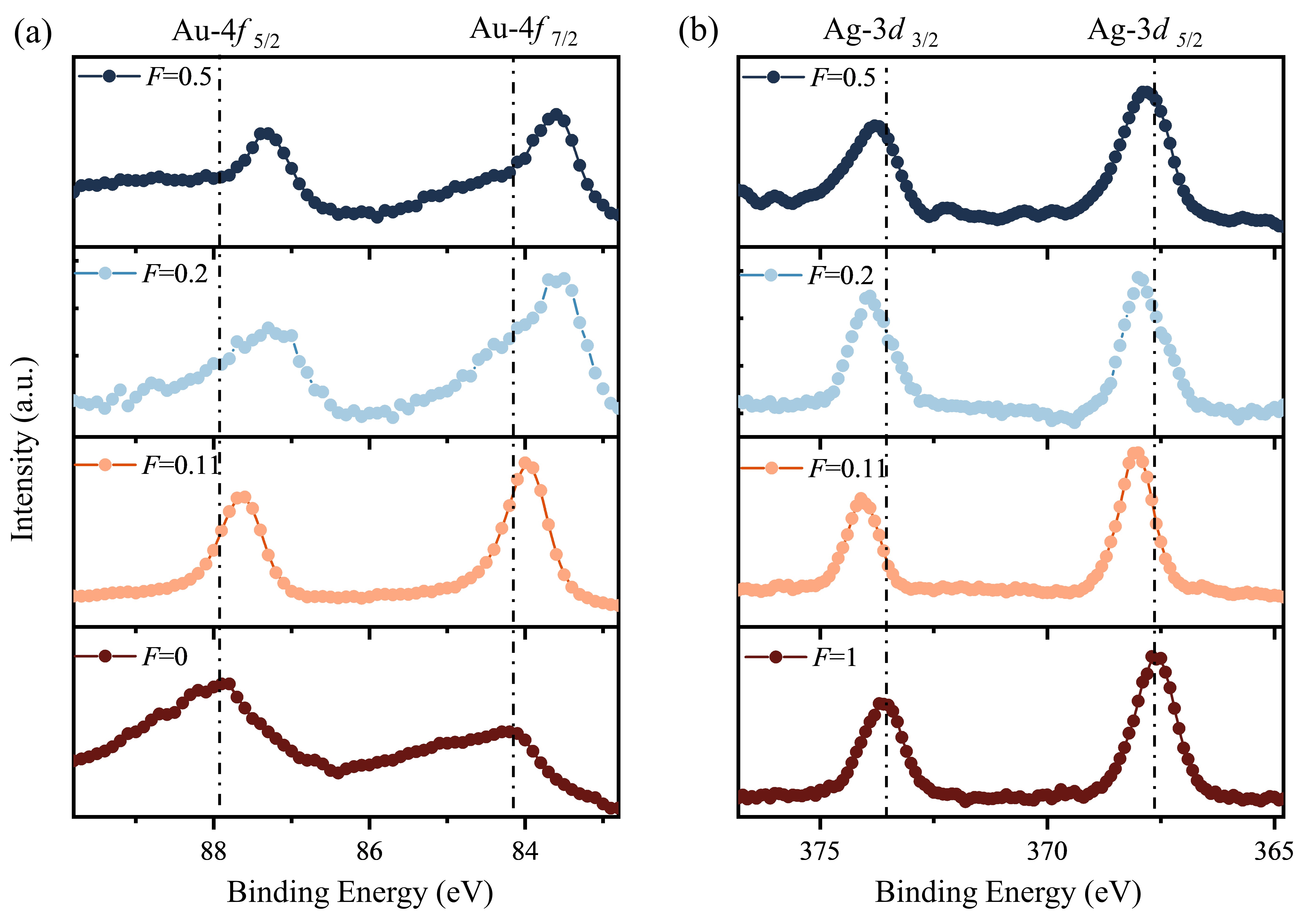}
 \caption{\textbf{ X-Ray Photoelectron Spectroscopy (XPS) spectrum of the Ag@Au nanohybrid:} \textcolor{black}{High-resolution peak  for (a) Au $4f$, and (b) Ag $3d$. The $4f$ peaks of Au shift toward low binding energy, indicating \textit{n}-type doping in the Au-matrix. At the same time, the $3d$ peaks of Ag shift towards higher binding energy, which indicates that Ag is losing electrons to Au. All the peaks are normalized with respect to the C=C peak at $284.5$~eV which is used as a charge correction reference.}}  
  \label{Extended Data Fig.7}
\end{figure*}
\begin{figure*}[t!]
\centering
  \includegraphics[height=6.2cm]{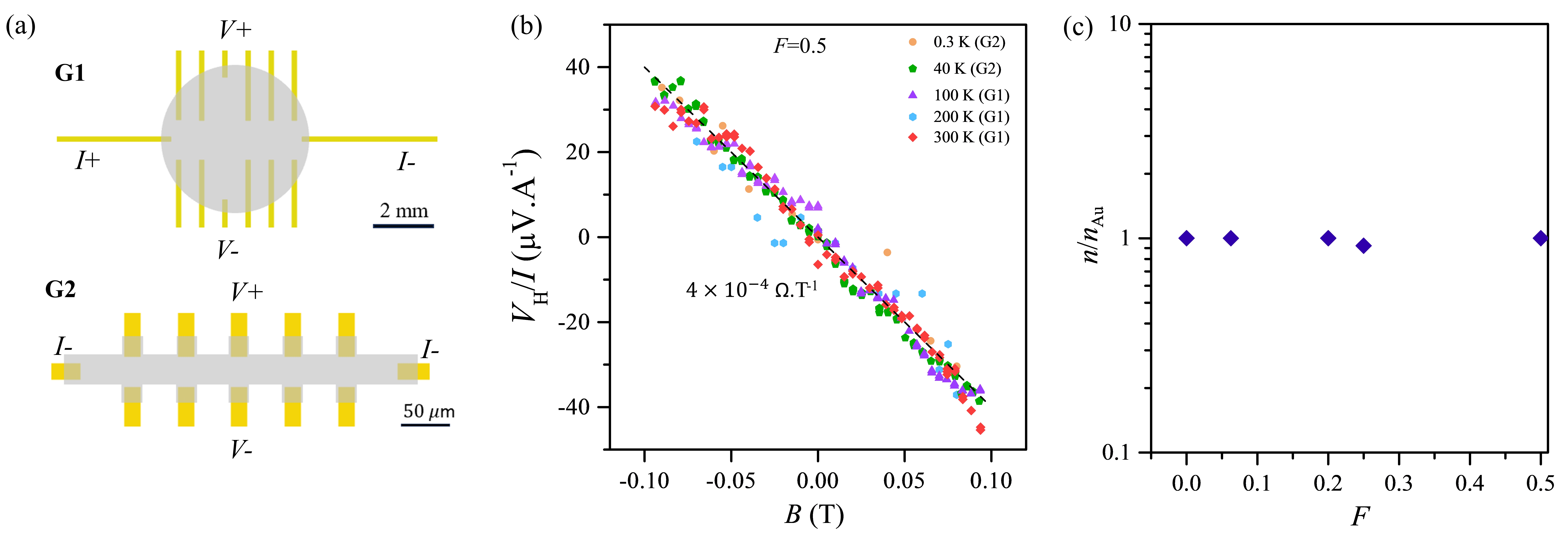}
 \caption{\textbf{Hall Measurements:} \textcolor{black}{(a) Top and bottom panels show the schematics of two diferrent lead geometries (G1 and G2) used for Hall measurements. Yellow lines show the depositd Cr/Au electrodes and the light grey region is the dropcast film. $I+,I-$ and $V+,V-$ indicate the current and voltage probes respectively. (b) Low field hall measurements at different temperatures ($T$) varying from $0.3-300$~K illustrated with a typical film with Ag fraction $F=V_{\mathrm{Ag}}/(V_{\mathrm{Ag}}+V_{\mathrm{Au}})=0.5$.}  Hall resistance ($V_\mathrm{H}/I$, $V_{\mathrm{H}}$, and $I$ are the measured hall voltage and current respectively)  is plotted as a function of magnetic field ($B$) at different temperatures ($T$). The dotted line indicates the expected linear variation of the Hall resistance as a function of $B$, which was the case in all nanoparticle films measured during the course of this experiment.
(c) The electron density $n~(\sim 10^{28}~\mathrm{m}^{-3})$ for different $F$ normalized to the magnitude of that ($n_{\mathrm{Au}}$) obtained in bare AuNP film. This indicates that the electron density in the Ag@Au hybrids is the same to that of bare Au, irrespective of Ag fraction $F$.}

  \label{Extended Data Fig.8}
\end{figure*}

\section*{C\MakeLowercase{omputational details}}
\textcolor{black}{The theoretical calculations for the charge transfer were done on a  2D `toy model' where a periodic array of $(4\times4)$ clusters of ``Ag'' sites are surrounded by 48 ``Au'' sites forming a superlattice of $(8\times8)$ 2D supercells. The calculations were carried out using the model Hamiltonian described in Supplementary Information Section VII. The relative ratio of the number of atoms for Ag:Au has been taken to be 1:3 to resemble a typical fraction $F=0.25$ of the Ag@Au core@shell nanohybrid. Due to the mismatch of the local potential seen by the conduction electrons localized in Wannier orbitals at the Ag and Au sites, as shown in Fig.~3c, and the long-range Coulomb interactions between electrons occupying these Wannier orbitals, there is charge transfer at each site, indicated by the excess electron occupancy $\delta n$. The amount of charge transferred can be tuned by varying the onsite potential difference between Au and Ag atoms $\epsilon_0$. As shown in Fig.~3c of the text, Au and Ag atoms become electron and hole-doped, respectively, for most positive values of $\epsilon_0$, which is also consistent with the X-ray photoelectron spectroscopy (XPS) data. \\
The parameters used in our study are as follows: the nearest neighbor distance, $d_1$, within the square lattice considered is set to $4.10$ \AA, corresponding to the lattice spacing in FCC-Ag/Au. The nearest neighbor hopping between all the atomic sites is set at $1$~eV, and the hopping decay factor, $\xi_0$, is adjusted to yield a second-nearest neighbor hopping of $0.5$~eV. The on-site Coulomb interaction energy is fixed at $U_0 = 2$~eV, and $V_0$, which controls the strength of the long range coulomb interactions, is chosen such that  $V_0/d_1 = 1$ eV. The difference in work function between Au and Ag atoms is incorporated into the on-site potentials, with $ (\epsilon_j^{\text{Ag}} - \epsilon_j^{\text{Au}})$ ranging from $0.1$ eV to $1.5$ eV in our calculations. The charge density obtained from solving the self-consistent mean field equations at $ (\epsilon_j^{\text{Ag}} - \epsilon_j^{\text{Au}}) = 1.5$ eV is presented in Fig. ~3d of the main manuscript. More detailed figures are provided in the Supplementary Information Section VII.\\
Phonons are computed with spring constants of $1.0$ for nearest neighbor atoms and $0.5$ for next-nearest neighbors, in arbitrary units. Electronic and phononic band spectra are calculated on $(16 \times 16)$ $\mathbf{k,q}$ grids within the Brillouin zone of the supercell, and are used to determine the electron-phonon coupling strength $(\lambda^{\mathrm{(calc)}})$ using the `double delta' approximation.}

\clearpage
\end{document}